\documentclass[pre, twocolumn, floatfix, superscriptaddress,nofootinbib,longbibliography]{revtex4-2}
\usepackage[utf8]{inputenc}
\usepackage{graphicx}
\usepackage{hyperref}
\usepackage{amssymb}
\usepackage{amsmath}
\usepackage{microtype}
\usepackage[table,dvipsnames]{xcolor}
\usepackage{physics}
\usepackage{dsfont}
\usepackage[export]{adjustbox}
\usepackage{booktabs}
\usepackage{multirow}
\usepackage{tabularx}
\usepackage{soul}
\usepackage{setspace}

\hypersetup{
    colorlinks = true,
    linkcolor  = blue,
    citecolor  = blue,
    urlcolor   = blue,
}
\renewcommand{\vec}[1]{\mathbf{#1}}

\renewcommand{\thesection}{\arabic{section}}
\renewcommand{\thesubsection}{\thesection.\arabic{subsection}}
\renewcommand{\thesubsubsection}{\thesubsection.\arabic{subsubsection}}
\makeatletter
\renewcommand{\p@subsection}{}
\renewcommand{\p@subsubsection}{}
\makeatother

\makeatletter
\g@addto@macro\bfseries{\boldmath}
\makeatother

\def\U#1{\text{U}(#1)}
\DeclareMathOperator{\sign}{sgn}

\begin{document}

\title{Hidden quasiconservation laws in fracton hydrodynamics}
\author{Oliver Hart}
\affiliation{Department of Physics and Center for Theory of Quantum Matter, University of Colorado, Boulder, Colorado 80309 USA}
\affiliation{T.C.M.~Group, Cavendish Laboratory, JJ Thomson Avenue, Cambridge CB3 0HE, United Kingdom}
\author{Andrew Lucas}
\email{andrew.j.lucas@colorado.edu}
\affiliation{Department of Physics and Center for Theory of Quantum Matter, University of Colorado, Boulder, Colorado 80309 USA}
\author{Rahul Nandkishore}
\affiliation{Department of Physics and Center for Theory of Quantum Matter, University of Colorado, Boulder, Colorado 80309 USA}
\date{October 15, 2021}

\begin{abstract}
    \setstretch{1.1}
    We show that the simplest universality classes of fracton hydrodynamics in more than one spatial dimension, including isotropic theories of charge and dipole conservation, can exhibit hidden ``quasiconservation laws", in which certain higher multipole moments can only decay due to dangerously irrelevant corrections to hydrodynamics.  We present two simple examples of this phenomenon.  Firstly, an isotropic dipole-conserving fluid in the infinite plane conserves an infinite number of ``harmonic multipole charges" within linear response; we calculate the decay or growth of these charges due to dangerously irrelevant nonlinearities. Secondly, we consider a model with $xy$ and $x^2-y^2$ quadrupole conservation, in addition to dipole conservation, which is described by isotropic fourth-order subdiffusion, yet has dangerously irrelevant sixth-order corrections necessary to relax the harmonic multipole charges.  We confirm our predictions for the anomalously slow decay of the harmonic conserved charges in each setting by using numerical simulations, both of the nonlinear hydrodynamic differential equations, and in quantum automaton circuits on a square lattice.
\end{abstract}

\maketitle


\section{Introduction}

Quantum dynamics constrained to conserve {\it multipole moments} of charge can exhibit striking phenomena that are only beginning to be understood. For example, locally generated quantum dynamics subject to conservation of charge and dipole moment can robustly break ergodicity~\cite{PPN, KhemaniShattering, SalaFragmentation}, with Hilbert space `shattering' into exponentially many dynamically disconnected `Krylov' subsectors, including a subspace exponentially large in system size within which the dynamics is exactly localized. This occurs in arbitrary spatial dimensions~\cite{KhemaniShattering} and has been the subject of intensive theoretical study~\cite{moudgalyaprem, SLIOMs, moudgalyamotrunich, khudorozhkov2021hilbert}. Thermalization in such `shattered' Hilbert spaces occurs not with respect to a symmetry sector, as is usually the case, but with respect to a Krylov subsector~\cite{moudgalyaprem}, and occurs more slowly than diffusion~\cite{IaconisSubsystem2019}. The universal, long wavelength description of such thermalization is given by `fracton hydrodynamics~\cite{fractonhydro},' and includes an infinitely large set of new hydrodynamic universality classes that are themselves arousing intense interest in the theory community~\cite{FeldmeierAnomalous2020, Morningstar, IaconisAnyDimension2021, Glorioso,Doshi:2020jso,Grosvenor:2021rrt}. 

We emphasize that such `multipolar' conservation laws are not just a theoretical curiosity. Indeed, they were predicted to occur~\cite{KhemaniShattering}, and have been observed experimentally~\cite{Bakr, Scherg2021observing,Kohlert2021experimental},
in tilted Fermi--Hubbard models in optical lattices. They have also been realized in systems of superconducting qubits~\cite{Guo}. Additionally,  `fracton hydrodynamics' constitutes the long wavelength theory for two-dimensional (2D) charged fluids in a magnetic field~\cite{fractonhydro}, and for certain dynamical universality classes of quantum magnets~\cite{Glorioso}. Finally, such `multipolar' conservation laws naturally arise~\cite{PretkoSubdimensional2017} in `fracton' phases of quantum matter: exotic quantum spin liquids where the elementary excitations have constrained mobility \cite{ChamonQuantumGlassiness2005, Haah2011, Vijay2015, VijayTopoOrder2016, PretkoSubdimensional2017, NandkishoreFractonsAnnurev}. 

This paper addresses a peculiar feature of the simplest fracton hydrodynamic theories, which appears to have been overlooked in the existing literature: the existence of (infinitely many) ``hidden quasiconservation laws": quantities that are not truly conserved, but which cannot decay within linear response. These quasiconservation laws both reveal an unexpected richness to the theoretical description of the problem, and could be directly accessible in near-term experiments, e.g., in optical lattices. 

\begin{figure}
    \centering
    \includegraphics[width=\linewidth]{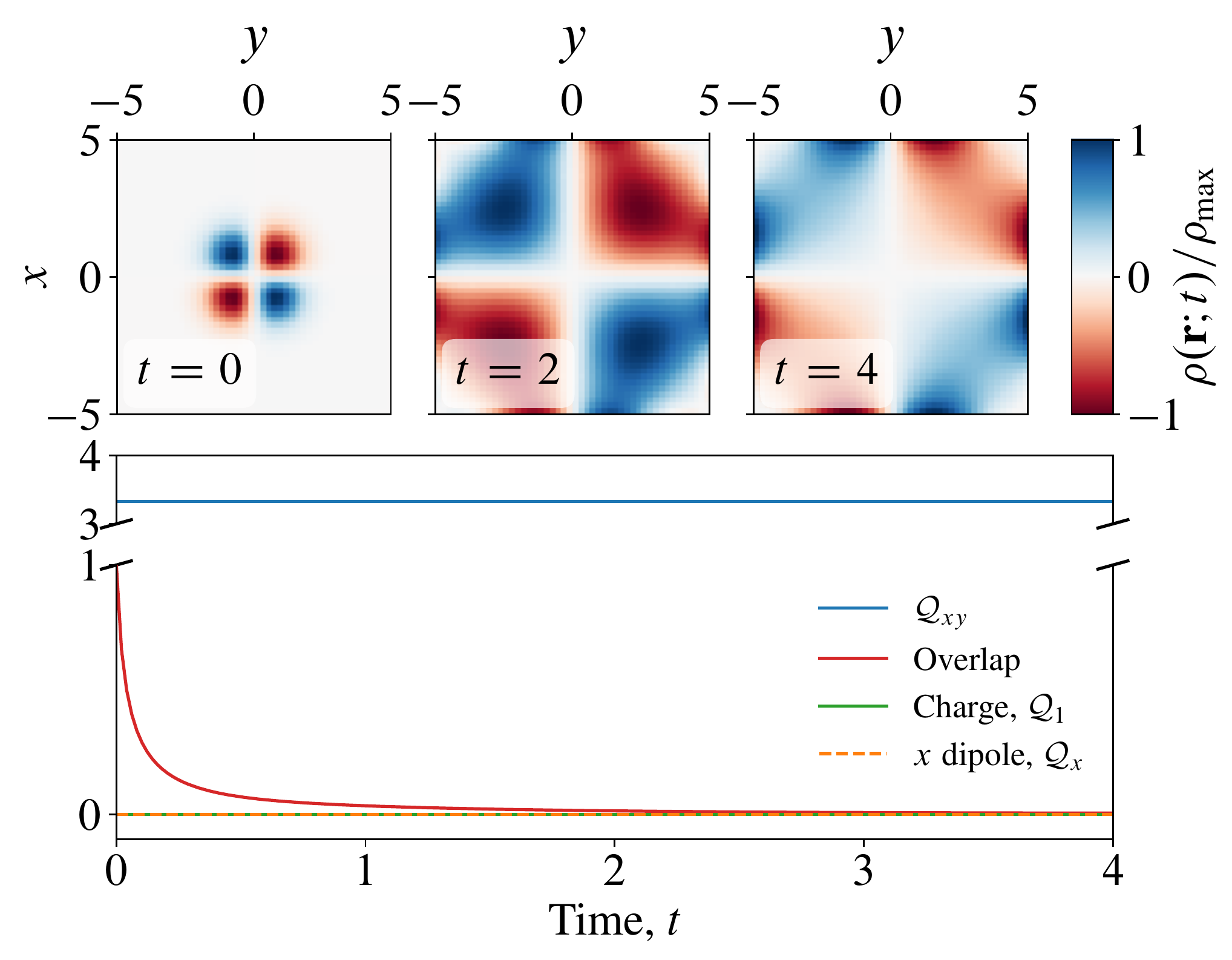}
    \caption{Solution of the isotropic, linear, dipole-conserving PDE $\partial_t \rho + D_0 \nabla^4 \rho = 0$ subject to boundary conditions conserving total charge and dipole moment: $\nabla^2 \rho = \partial_\vec{n} \nabla^2 \rho = 0$, with $\partial_\vec{n}$ the normal derivative. Observe that $\mathcal{Q}_{xy} = \int \mathrm{d}^2\mathbf{r} \, xy \rho $ is also conserved exactly. In fact, all harmonic functions $f(x, y)$ [$\nabla^2 f = 0$] give rise to a conserved charge $\mathcal{Q}_f = \int \mathrm{d}^2 \mathbf{r} \, f \rho$. We initialise the system with the density profile $\rho_0(\vec{r}) \propto xy \exp(-r^2/2\sigma^2)$, and plot the moments $f \in \{1, x, xy\}$ and the ``overlap'' with the initial state, $\int \mathrm{d}^2\mathbf{r} \,  \rho_0 \rho$ (normalised by its $t=0$ value). The density profiles $\rho(\vec{r}; t)$ in the top panels are normalised by their maximum values. The PDE is solved on a lattice with $N^2=48^2$ sites, and $\sigma \simeq 0.42$.} \label{fig:cover_image}
\end{figure}

We illustrate the quasiconservation laws with a simple example.  Consider the linear response theory of fourth-order subdiffusion, which arises due to dipole conservation~\cite{PretkoGeneralizedEM,fractonhydro}:
\begin{equation}
    \partial_t \rho = - D_0 (\nabla^2)^2 \rho \, ; \label{eq:introk4}
\end{equation}
here $\rho$ denotes the charge density and $D_0>0$ is a phenomenological constant.  All nonlinearities are irrelevant in any spatial dimension~\cite{fractonhydro}, and thus we do not consider them here.  Suppose that we solve this equation in an infinite $d$-dimensional volume, and ask what multipole moments of charge are conserved?  Namely, if we write
\begin{equation}
    \frac{\mathrm{d}}{\mathrm{d}t} \int \mathrm{d}^d\vec{r} \; f(\vec{r})\rho(\vec{r}; t) = 0 \, , \label{eq:ddtf}
\end{equation}
for which functions $f(\vec{r})$ is this equation true if $\rho$ solves Eq.~\eqref{eq:introk4}? After a simple calculation (spelled out in detail in the bulk of this paper), one finds that 
\begin{equation}
    \nabla^2 f = 0 \label{eq:harmonic}
\end{equation}
is already sufficient to satisfy (\ref{eq:ddtf}) in the infinite plane (with the correct number of boundary conditions at infinity).  In fact, even in certain finite domains, one can choose boundary conditions such that (\ref{eq:ddtf}) is satisfied whenever (\ref{eq:harmonic}) holds. 
In $d>1$, Eq.~\eqref{eq:harmonic} is satisfied by \emph{infinitely many $f$}: the harmonic functions. A simple example in $d=2$ is $f=xy$. These conservation laws were {\it not} built in explicitly, unlike charge and dipole conservation, but infinitely many of them seem to emerge all the same. Numerical simulations of~\eqref{eq:introk4}, presented in Fig.~\ref{fig:cover_image}, confirm that even in a domain of finite size with seemingly mundane boundary conditions, hidden conservation laws emerge within the linearized hydrodynamic theory.

The goal of this paper is to deduce what ultimately happens to the harmonic multipole charges
\begin{equation}
    {
    \mathcal{Q}_f \equiv \int \mathrm{d}^d \vec{r} \;
    f(\vec{r}) \rho(\vec{r}; t)
    \, ,}
    \label{eqn:harmonic-charge-definition}
\end{equation}
in a genuine charge- and dipole-conserving system, where there is certainly no true conservation law for an infinite number of $f$.  We will discuss two resolutions and confirm each with numerical simulations.

The first possible resolution, which is relevant for our cartoon scenario in the infinite isotropic continuum described above, is that formally irrelevant nonlinear corrections to the subdiffusion equation allow partial relaxation of harmonic multipole charges.  We will analytically predict algebraic-in-time decay or growth of these harmonic charges and confirm these expectations via numerical simulations. We note, however, that the irrelevant non-linearities decay fast enough that the charges saturate to a non-zero (but non-universal) value, i.e., in the infinite isotropic continuum, some fraction of the harmonic function charge does not decay. We then consider fracton hydrodynamics \emph{with boundaries}. We analytically argue that, even for an isotropic continuum theory, the combination of non-linearities and boundaries allows all harmonic function charges to relax to zero in the long time limit. Furthermore, we expect this relaxation to be exponential in time on the longest timescales, although charge density will need to `subdiffuse' to the boundary for the harmonic charges to relax efficiently.

It is important to note that most known physical realizations of fracton hydrodynamics occur not in the isotropic continuum, but rather on a {\it lattice}. In a lattice model, there will arise lattice anisotropies, which can themselves allow the harmonic charges to acquire dynamics, both in the bulk and on the boundary. If the most relevant term in fracton hydrodynamics is isotropic, then the relaxation will come from higher derivative, dangerously irrelevant, terms exhibiting lattice anisotropy. In this case, the harmonic multipole charges will relax more slowly (i.e., more subdiffusively) than the simple power counting from fracton hydrodynamics would suggest. This represents another example of `UV-IR mixing'~\cite{BulmashBarkeshli2018, Gromov2019, Hart2021TypeII, Schmit}, in that microscopic lattice scales show up in the long wavelength hydrodynamic description.

Our treatment in the continuum will be based on analytical and numerical solutions of the partial differential equations of fracton hydrodynamics. On the lattice, we will supplement these analyses with automaton Monte Carlo simulations in the manner of Ref.~\cite{IaconisSubsystem2019}. We will also illustrate the above principles in the specific context of a fracton fluid inspired by the \U1 generalisation of Haah's code \cite{BulmashBarkeshli2018,Gromov2019,Gromov2020duality,WesleiChamon2021,Hart2021TypeII}. We conclude with a discussion of the implications of our results.


\section{2D automaton circuits}

To test our predictions numerically on the lattice, we make use of cellular automaton quantum circuits~\cite{GopalakrishnanAutomata2018,AlbaRule54,IaconisSubsystem2019,IaconisPRXComplexity}.
For a more in-depth overview of the properties of automaton circuits, we refer the reader to, e.g., Refs.~\cite{IaconisSubsystem2019,FeldmeierAnomalous2020,IaconisAnyDimension2021,IaconisPRXComplexity}.

Cellular automaton dynamics is a type of discrete unitary time evolution that does not generate entanglement in a particular, privileged (local) basis.
That is, an automaton gate $\hat{U}$, when acting on a state $\ket{m}$ belonging to this privileged basis, simply permutes the basis states, and returns another state belonging to the same basis, up to a phase~\cite{IaconisSubsystem2019,IaconisPRXComplexity}.
For a $D$-dimensional basis, this operation can be represented by $\hat{U}\ket{m} = e^{i\theta_m} \ket{\pi_D(m)}$, where $\pi_D \in S_D$ is an element of the permutation group on $D$ elements.
While no real-space entanglement is generated when acting with $\hat{U}$ upon a basis state $\ket{m}$,
the time evolution generated by repeated application of automaton gates generates volume law entanglement when acting upon a generic initial state $\ket{\psi} = \sum_{m=1}^D c_m \ket{m}$ ~\cite{IaconisPRXComplexity}.
Indeed, while the evolution of a privileged basis state $\ket{m}$ can be performed entirely classically, generic automaton dynamics gives rise to chaotic quantum dynamics~\cite{IaconisPRXComplexity}, being able to reproduce many of the properties of the more general Haar random circuits.

\begin{figure}
    \centering
    \includegraphics[width=\linewidth]{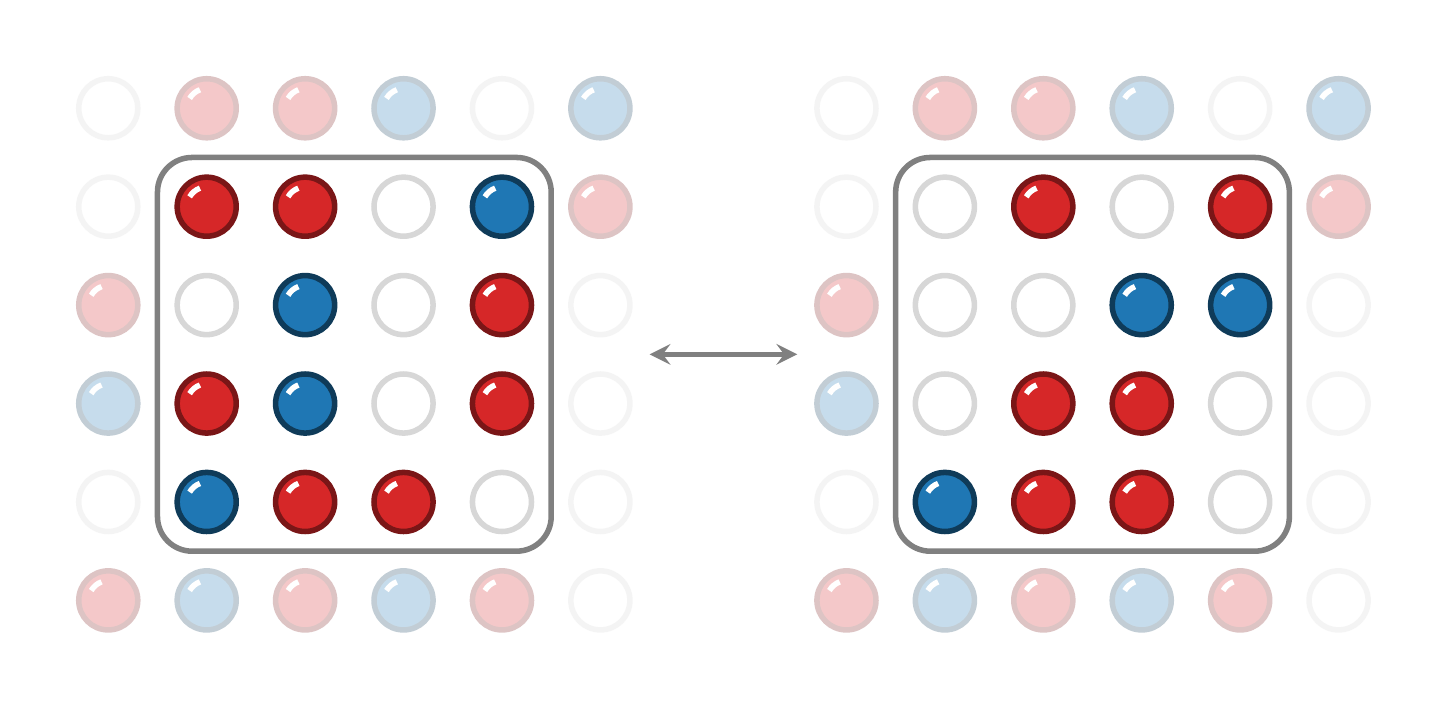}
    \caption{Illustration of the $4 \times 4$ dipole-conserving gates that are employed in the automaton circuits. The blue (red) circles indicate the presence of a positive (negative) charge, i.e., $S_i^z=1(-1)$, while the empty circles denote unoccupied sites, i.e., $S_i^z=0$. Both the left and the right charge configurations belong to the same charge $\sum_i S_i^z =-3$, and dipole $\sum_i \vec{r}_i S_i^z = (-5, -4)$ sector (with the origin at the bottom left of the gate, and the lattice spacing $a=1$).}
    \label{fig:dipole-conserving-gates}
\end{figure}

Throughout the manuscript the privileged basis will correspond to local spin-1 tensor product states of the form $\ket{m}=\otimes_i \ket{m_i}$ with $\ket{m_i} \in \{\ket{+}, \ket{0}, \ket{-}\}$ the eigenstates of $\hat{S}_i^z$ on site $i$.
We will refer to the states $\ket{\pm}$ as hosting a charge $q=\pm 1$.
Therefore, in each discrete time update of the automaton evolution, an $\hat{S}_i^z$ product state is mapped onto another $\hat{S}_i^z$ product state.
Gates of the form $\prod_{j}\hat{U}_{j,t}$ are applied sequentially to the system at each time step $t$.
The local gates $\hat{U}_{j,t}$, which we take to be randomly and uniformly distributed over space, permute the spin states within a fixed symmetry sector set by the global conservation laws, as indicated schematically in Fig.~\ref{fig:dipole-conserving-gates} for gates of size $G_x \times G_y = 4 \times 4$ in two dimensions\footnote{We define one unit of time as $N/G$ random gate applications, where $N$ is the total number of spins, and $G=G_xG_y$ is the number of spins acted upon by each gate.}.

\subsection{Decay of hydrodynamic modes}

\begin{figure}[t]
    \centering
    \includegraphics[width=\linewidth]{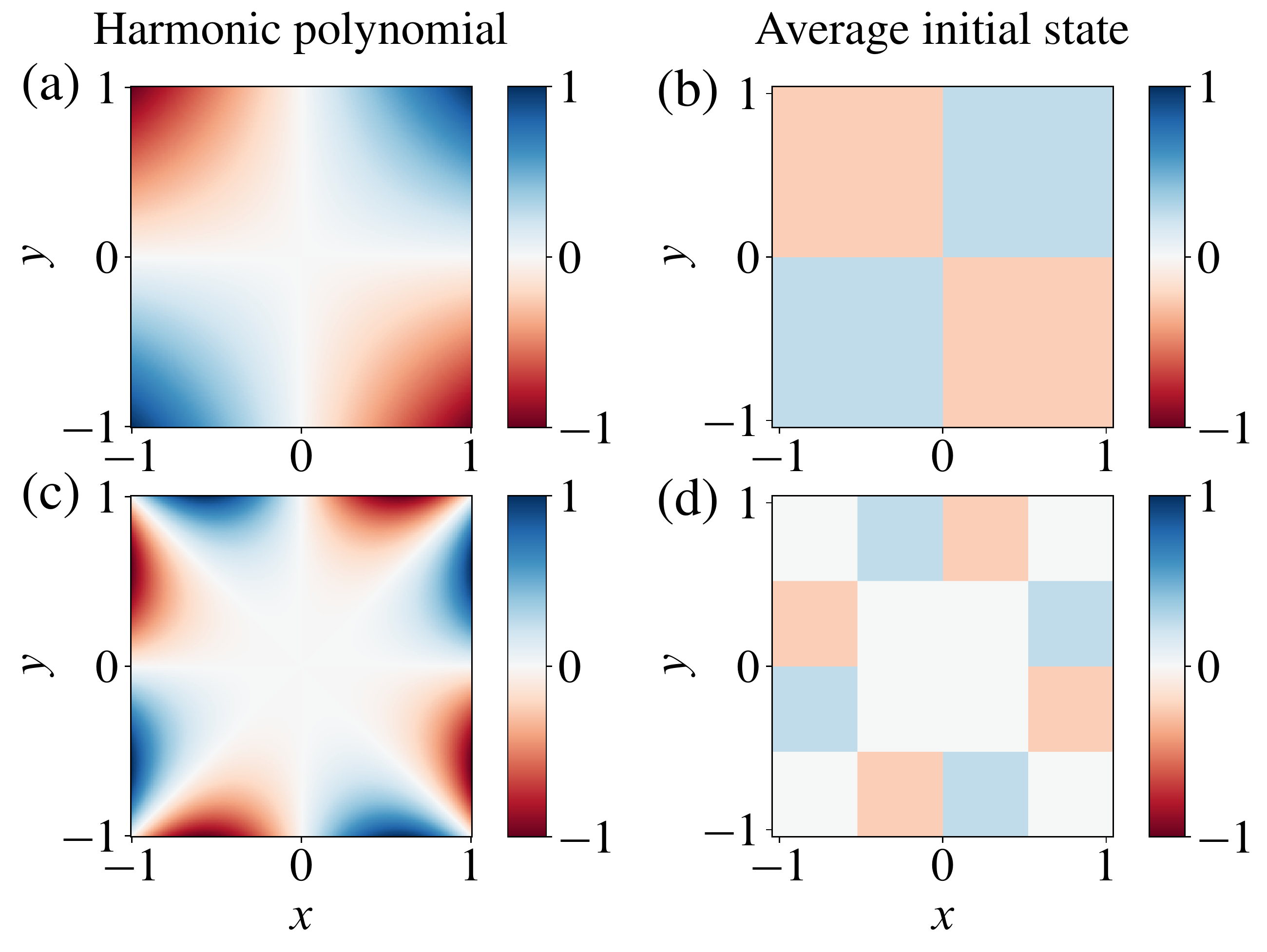}
    \caption{Left column: Illustration of the polynomials $x y$ (top) and $x^3 y - x y^3$ (bottom). Right column: The corresponding \emph{average} initial states used for the automaton circuits: the charge on each site is drawn from a weakly biased infinite temperature distribution, as described in the main text. The initial states in the right column therefore have strong overlap with the corresponding harmonic polynomials depicted in the left column.}
    \label{fig:automaton-states}
\end{figure}

One tool that we use to test the predictions of hydrodynamics is probing the relaxation of specific initial charge distributions.
In particular, we initialise the system at time $t=0$ in a spin-1 product state in the computational basis, $\ket{\psi_z(0)} = \ket{\{ S_i^z \}}$.
The state $\ket{\psi_z(0)}$ is chosen to have particularly strong overlap with one of the putative hydrodynamic modes (and/or the relevant harmonic charges).
The time evolution generated thereafter by the automaton circuit is followed, and the ``overlap'' with the initial state can be used to diagnose how the system relaxes towards equilibrium.
The initial charge configurations that we utilise (depicted in Fig.~\ref{fig:automaton-states}) have net zero charge and dipole moment\footnote{This statement is true only on average; in a system of finite size, there is statistical uncertainty in the total charge and dipole moment which vanishes in the thermodynamic limit. Of course, the charge and dipole in the statistically sampled initial conditions are exactly conserved within each realization.}, so the equilibrium state corresponds, in the absence of shattering\footnote{The neglect of shattering can be justified by assuming that the gates are large enough that the shattering is weak in the sense of Refs.~\cite{KhemaniShattering, SalaFragmentation}, i.e., that almost all of Hilbert space thermalizes, and also that the initial condition does not lie in the (exponentially large but measure zero) `localized' portion of Hilbert space.}, to a featureless infinite temperature distribution of $\{S_i^z\}$ within a fixed symmetry sector.

In the case of a ``quadrupole'' initial state [i.e., a state with a nontrivial quadrupole moment, as depicted in Fig.~\ref{fig:automaton-states}(b)],
for example, the system is split into four quadrants. The top right and bottom left (top left and bottom right) quadrants are assigned a net positive (negative) charge. In the net positive regions, this is accomplished by drawing each $S_j^z$ from the biased probability distribution $P(S_i^z)$
\begin{equation}
    P(1) = \frac{1+2\bar{\rho}}{3} \, , \quad P(0) = P(-1) = \frac{1-\bar{\rho}}{3}
    \, .
    \label{eqn:on-site-probability-dist}
\end{equation}
Implying that $\overline{\langle\psi_z(0) |\hat{S}_i^z|\psi_z(0) \rangle} = \bar{\rho} > 0$, where the angled brackets denote the quantum expectation value, and the overline represents an ensemble average over the probability distribution~\eqref{eqn:on-site-probability-dist} on each site, i.e., $\overline{A}=\sum_{\{S_i^z\}} A(\{S_i^z\}) \prod_j P(S_j^z) $.
The average density $\bar{\rho}$ is chosen to be small, $\bar{\rho} \ll 1$, both to avoid obstructions in the path to equilibrium in the form of strong shattering~\cite{KhemaniShattering,SalaFragmentation, Morningstar}, and to minimise any short-time transient dynamics before the hydrodynamic description sets in.
In the negatively charged quadrants, the probability distribution~\eqref{eqn:on-site-probability-dist} is reversed (i.e.,
$\bar{\rho} \to - \bar{\rho}$ in these regions), thereby favouring negative $S_i^z$
eigenvalues.

The method of initialisation described above can alternatively be considered as a projection of the initial state
\begin{equation}
    \ket{\Psi(0)} \propto  \bigotimes_{i=1}^N \left[ (1+h_{i+}) \ket{+} + \ket{0} + (1+h_{i-})\ket{-} \right]
    \, ,
    \label{eqn:spin-1-initstate}
\end{equation}
onto the computational basis $\ket{\{ S_i^z \} }$,
where the parameters $h_{i\pm}$ are related to the average density $\bar{\rho}$ via $h_{i+(-)}=\sqrt{\frac{1+2\bar{\rho}}{1-\bar{\rho}}}-1$ and $h_{i-(+)}=0$ for $i$ belonging to the positive (negative) quadrant.
Each state $\ket{\{ S_i^z \} }$ appearing in the superposition is then evolved separately according to the automaton dynamics.
This leads to the relationship\footnote{The approximate equality becomes an identity if (i) the average over the $(2S+1)^N$ initial product states is performed exactly, and (ii) each product state appearing in the decomposition is evolved according to an identical set of automaton gates.} $\langle \Psi(t) | \hat{S}_i^z  | \Psi(t) \rangle \simeq \overline{\langle\psi_z(t) |\hat{S}_i^z|\psi_z(t) \rangle}$ for normalised states $\ket{\Psi(0)}$ and $\ket{\psi_z(0)}$.
The emergent averaging that occurs for initial states such as~\eqref{eqn:spin-1-initstate} is reminiscent of disorder-free localised systems~\cite{Smith2017DisorderFree,Smith2019Thesis}, in which an extensive number of local symmetries lead to an emergent disorder average for certain translationally invariant initial states.

The ``overlap'' with the initial charge configuration can be quantified by the imbalance $\mathcal{I}(t)$ between charge in the positively versus negatively charged regions 
\begin{subequations}\label{eqn:imbalance-definition}
\begin{align}
    \mathcal{I}(t) &= \frac{1}{N\bar{\rho}}\left[
    \sum_{i \in R_+} \overline{\langle\hat{S}_i^z(t)\rangle}
    - \sum_{i \in R_-} \overline{\langle\hat{S}_i^z(t)\rangle}
    \right] \\
    &\equiv \frac{1}{N\bar{\rho}^2} \sum_{i=1}^N \overline{\langle \hat{S}_i^z(t) \rangle}
    \:\:
    \overline{\langle \hat{S}_i^z(0) \rangle}
    \, ,
\end{align}
\end{subequations}
where the region $R_{+(-)}$ is the set of spins initialised with net positive (negative) charge density.
It follows from the definition~\eqref{eqn:imbalance-definition} that $\mathcal{I}(0)= 1$.
The imbalance will subsequently decay from this maximal value at $t=0$. The asymptotic decay of $\mathcal{I}(t)$ is governed by the slowest hydrodynamic mode with which the initial condition has nonzero overlap.


\subsection{Spin correlation function}

We also measure the infinite temperature spin correlation function in real space:
\begin{equation}
    C_z(\vec{r}_i - \vec{r}_j; t) = \overline{\langle \hat{S}_i^z(t) \hat{S}_j^z(0) \rangle}
    \, .
\end{equation}
Here, the overline denotes an average over $\ket{\{S_i^z \} }$ product states according an infinite temperature probability distribution, i.e., Eq.~\eqref{eqn:on-site-probability-dist} with zero net charge density $\bar{\rho}=0$ on all sites.
 The time dependence of $C_z(\vec{0}; t)$ quantifies the return probability; if the charge density spreads subdiffusively ($r \sim t^{1/4}$ for dipole conserving systems) then conservation of charge implies that $C_z(\vec{0}; t) \sim t^{-d/4}$.
This prediction has been verified in $d=1$ in
Refs.~\cite{IaconisSubsystem2019,FeldmeierAnomalous2020,IaconisAnyDimension2021}.

In spatial dimensions $d \geq 2$, lattice anisotropy plays an important role in determining the spatial dependence of $C_z(\vec{r}; t)$ for fixed $t$.
In conventional diffusive hydrodynamics, diffusion is isotropic because there are no nontrivial second rank tensors that are invariant under the point group symmetry of the lattice.
Meanwhile, for \emph{fracton} hydrodynamics, 
in which terms second order in spatial derivatives are removed by the additional conservation laws, there exist nontrivial higher rank tensors that are invariant under the lattice's point group~\cite{fractonhydro}.
The conserved charge density evolves according to the canonical equation of fracton hydrodynamics~\cite{PretkoGeneralizedEM,fractonhydro}
\begin{equation}
    \partial_t \rho + \partial_i \partial_j J^{ij} = 0
    \, .
    \label{eqn:fracton-continuity}
\end{equation}
To lowest order in spatial derivatives, the constitutive relation between dipole current $J^{ij}$ and the charge density $\rho$ takes the form $J^{ij} = D^{ijk\ell}\partial_k \partial_\ell \rho$~\cite{IaconisAnyDimension2021},
where the index structure of $D^{ijk\ell}$ is constrained by the point group  of the lattice.
A density modulation characterised by wave vector $\vec{k}$ will therefore decay as $\sim e^{-\Gamma(\vec{k}) t}$, where the decay rate $\Gamma(\vec{k}) = D^{ijk\ell} k_i k_j k_k k_\ell$ is in general anisotropic.

Putting the above ingredients together, the hydrodynamic prediction for the correlation function $C_z(\vec{r}; t)$ is
\begin{align}
    C_z(\vec{r}; t) &= \frac{1}{L^d} \sum_{\vec{k}, \vec{p}} e^{i\vec{k}\cdot \vec{r}} 
    \overline{\langle \hat{S}^z(\vec{k}, t) \hat{S}^z(\vec{p}, 0) \rangle} \notag  \\
    &\simeq \frac{1}{3}S(S+1) \int \frac{\mathrm{d}^d \vec{k}}{(2\pi)^d} e^{i\vec{k}\cdot \vec{r}} e^{-\Gamma(\vec{k}) t}
    \label{eqn:hydro-correlator}
\end{align}
where the $S$-dependent prefactor comes from the infinite temperature average $\overline{\langle \hat{S}_i^z \hat{S}_j^z \rangle}=\tfrac13 \delta_{ij} S(S+1)$, and we have assumed (discrete) translational invariance.
Equation~\eqref{eqn:hydro-correlator} holds for sufficiently long times and distances, i.e., in the hydrodynamic limit $|\vec{r}| \to \infty$ and $t \to \infty$ keeping $|\vec{r}|^4/t$ fixed. The spatial dependence of $C_z(\vec{r}; t)$ therefore provides access to important information about the constitutive relation $J^{ij}(\rho)$.

\setlength{\tabcolsep}{10pt}
\begin{table}[b]
\caption{The lowest-degree harmonic polynomials relevant to this work.}
\label{tab:polynomials}
\begin{tabularx}{\linewidth}{@{}clX@{}}
\arrayrulecolor{black}\toprule
Order, $n$         & Conserved quantity          & Polynomial    \\ \arrayrulecolor{black}\midrule
0                  & Charge                      & $1$                            \\ \arrayrulecolor{black!30}\midrule
\multirow{2}{*}{1} & \multirow{2}{*}{Dipole}     & $x$                          \\
                   &                             & $y$                          \\ \arrayrulecolor{black!30}\midrule
\multirow{2}{*}{2} & \multirow{2}{*}{Quadrupole} & $x^2 - y^2$                \\
                   &                             & $x y$                      \\ \arrayrulecolor{black!30}\midrule
\multirow{2}{*}{4} & \multirow{2}{*}{Four-pole}   & $x^4 - 6 x^2 y^2 + y^4$ \\
                   &                             & $x^3 y - x y^3$        \\ \arrayrulecolor{black}\bottomrule
\end{tabularx}
\end{table}


\section{Dynamics of harmonic function charges in the isotropic continuum}
\subsection{The infinite plane}
\label{sec:nonlinear-corrections}

Here we describe how the additional quasiconserved harmonic charges $\mathcal{Q}_f$ [defined in Eq.~\eqref{eqn:harmonic-charge-definition}] are given dynamics by the presence of nonlinearities in the constitutive relation between charge and dipole current.
In the isotropic case, the simplest nonlinear correction that we can include takes the form of a density-dependent subdiffusion constant:\footnote{In general, the first subleading term in the expansion is $D_1\rho$, not $D_2\rho^2$ (especially if we consider $\rho$ to be a perturbation about a finite density state).  However, we have explicitly checked that the scaling argument that we present in this paper is unchanged if one accounts for this linear correction, albeit becoming more cumbersome to write down.  This happens because a linear term $D_1\rho$ only leads to relaxation of $\mathcal{Q}_{xy}$ at second order in $D_1$, while $D_2\rho^2$ can relax $\mathcal{Q}_{xy}$ at first order.}
\begin{equation}
    J_{ij} = (D_0 + D_2\rho^2 + \ldots) \partial_i \partial_j \rho
    \, .
    \label{eqn:nonlinear-constitutive}
\end{equation}
To scrutinise the conservation of the lowest-degree harmonic charges, consider
the hydrodynamic equation for the evolution of the quadrupole density $x^\ell x^m \rho$ (note that $\ell$ and $m$ are indices, not exponents):
\begin{equation}
    \partial_t(x^\ell x^m \rho) + \partial_i J^{i\ell m}_\text{q} = -2J^{\ell m}
    \, ,
    \label{eqn:xy-continuity}
\end{equation}
where the quadrupole current $J_\text{q}^{i\ell m} = x^\ell x^m \partial_j J^{ij} - x^m J^{i\ell} - x^\ell J^{im}$. Equation~\eqref{eqn:xy-continuity}
is a special case of the general $n$-pole continuity equation presented in Appendix~\ref{sec:n-pole-continuity}, and
follows directly from the continuity equation for charge~\eqref{eqn:fracton-continuity}, noting that $\partial_i (x^\ell x^m \partial_j J^{ij}) = x^\ell x^m \partial_i \partial_j J^{ij} + x^\ell  \partial_j  J^{mj} + x^{m}  \partial_j J^{\ell j} = x^\ell x^m \partial_i \partial_j J^{ij} + \partial_i(x^m J^{i\ell} + x^\ell J^{i m}) - 2J^{\ell m}$.
Evidently, the quadrupole density $x^ix^j\rho$ is sourced by the dipole current tensor $J^{ij}$. 

In the linear theory, i.e., in the absence of $D_2$ and higher order corrections, the appropriate components of the dipole current tensor $J_{ij}$ can be written as a total derivative, e.g.,
$J_{xy}\propto \partial_x \partial_y \rho $ (and similarly for $J_{xx} - J_{yy}$).
However, as soon as nonlinearities are included, this relationship ceases to hold, and there exists a source term for the harmonic charges $\mathcal{Q}_{xy}$ and $\mathcal{Q}_{xx}-\mathcal{Q}_{yy}$ that cannot be rewritten as a total derivative.
The boundary terms that arise from integrating by parts also play an important role in systems of finite size, and will be discussed in further detail in the next section.

\begin{figure}
    \centering
    \includegraphics[width=\linewidth]{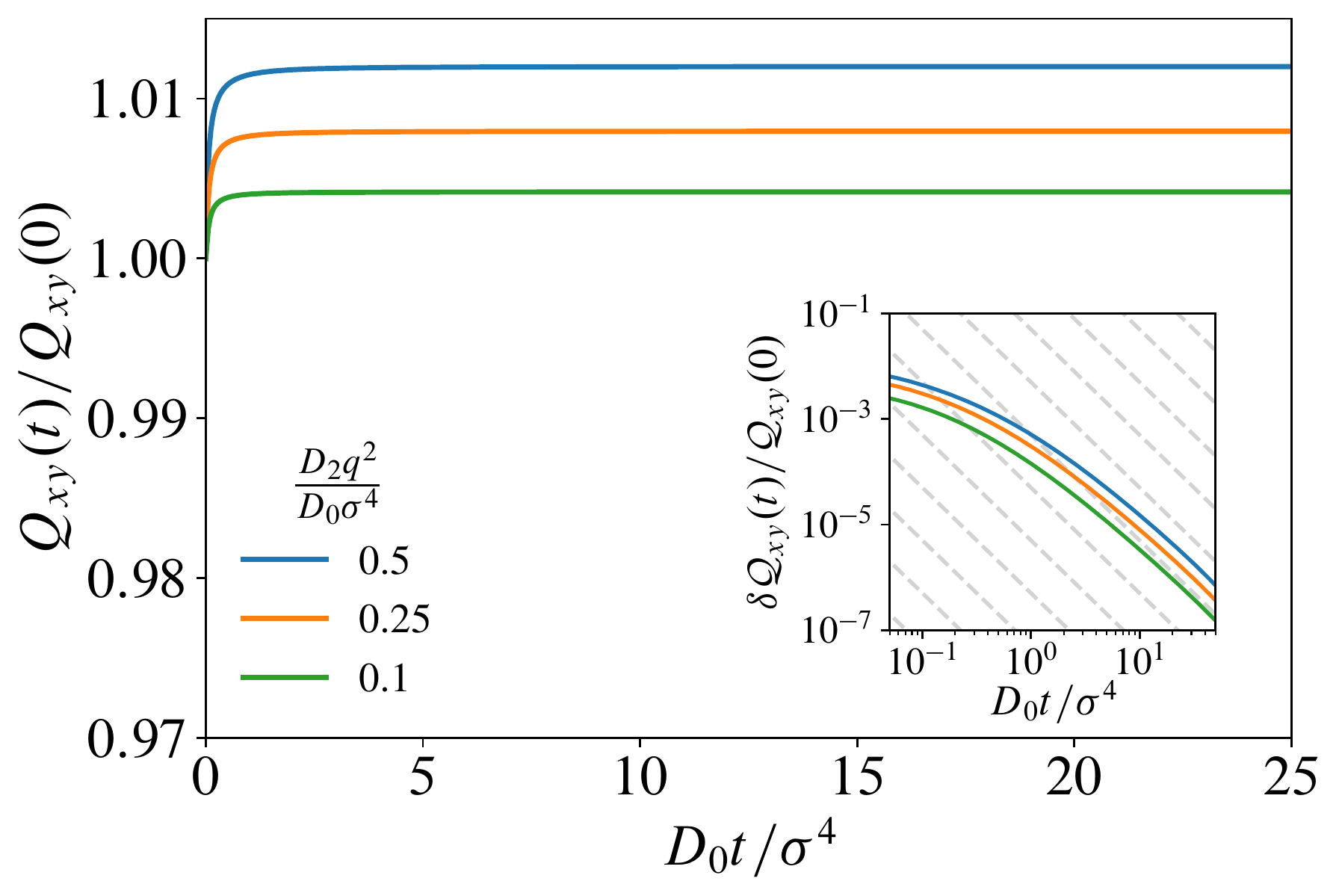}
    \caption{Dynamics of the $xy$ charge, $\mathcal{Q}_{xy}(t)$, in the presence of a nonlinear correction to the constitutive relation between charge and dipole current [Eq.~\eqref{eqn:nonlinear-constitutive}] for various $D_2q^2/(D_0 \sigma^4)$, a dimensionless parameter that quantifies the nonlinearity. The relaxation of the charge towards its asymptotic value, $ \mathcal{Q}_{xy}(\infty) - \mathcal{Q}_{xy}(t) \sim t^{-\alpha}$, is consistent with $\alpha=2$, as shown in the inset (the grey dashed lines have an exponent of $-2$). Note that the charge is not relaxed to zero; some fraction of the harmonic function charge truly does not decay in the long time limit, and the sign of $D_2$ dictates whether the charge grows or decays from its initial value. A system of linear size $L/\sigma\simeq 60$ is initialised with a charge density $\rho(\vec{r}; 0) = (q/\sigma^4)xye^{-r^2/2\sigma^2}$.}
    \label{fig:nonlinear-relxation}
\end{figure}

In the presence of such nonlinearities, the equation of motion for the $xy$ charge is
\begin{align}
    \frac{\mathrm{d}}{\mathrm{d}t} \mathcal{Q}_{xy}  &= -2\int \mathrm{d}^2\mathbf{r} \; J_{xy} + \text{boundary terms} \notag \\
    &= -2D_2 \int \mathrm{d}^2\mathbf{r} \; \rho^2 \partial_x\partial_y \rho + \text{boundary terms}
    \, . \label{eq:isotropiccontinuumrelaxation}
\end{align}
To calculate the dynamics of the $xy$ charge implied by~\eqref{eq:isotropiccontinuumrelaxation} within perturbation theory, we must first solve the linear problem.
The Green's function for the isotropic subdiffusion equation~\eqref{eq:introk4} in two dimensions is given by the Hankel transform of $e^{-D_0 k^4 t}$
\begin{equation}
    G(\vec{r}; t) = \int\frac{dk}{2\pi}\, k J_0(kr) e^{-D_0 k^4 t} = \frac{1}{(D_0t)^{1/2}}f\left( u \right)
    \, ,
    \label{eqn:dipole-conserving-Greens-function}
\end{equation}
where $u=r(D_0 t)^{-1/4}$.
The scaling function $f(u)$ can be represented explicitly in terms of hypergeometric functions (see Appendix~\ref{sec:linear-exact-solution}).
Equation~\eqref{eqn:dipole-conserving-Greens-function} describes how a unit $\delta$-distributed charge density spreads subdiffusively in time from its initial position.
However, such an initial condition has a trivial dipole and quadrupole moment (zero, if the origin corresponds to the initial position).
To obtain a nontrivial $xy$ charge, we should initialise the system in a charge configuration with nonzero quadrupole moment.
This may be accomplished by considering two oppositely orientated dipoles
\begin{equation}\label{eqn:quadrupole-charge-config}
    \rho_\text{q}(\vec{r}; 0) = \sum_{\substack{\sigma_x,\sigma_y \\ = \pm 1}} \sigma_x\sigma_yq \delta\left(x-\frac{a}{2}\sigma_x\right)\delta\left(y-\frac{a}{2}\sigma_y\right).
\end{equation}
Taking $a\to 0^+$ while keeping the quadrupole moment $Q = a^2 q$ fixed, the time dependence of the initial charge configuration~\eqref{eqn:quadrupole-charge-config} is given by
$\rho_\text{q}(\vec{r}; t) = Q \partial_x \partial_y G(\vec{r}; t)$.
In an analogous manner, to obtain nonzero overlap with higher-degree harmonic polynomials, such as those listed in Table~\ref{tab:polynomials}, one must use an initial charge distributions with nonzero higher multipole moments, whose time dependence is determined by higher-order derivatives of $G(\vec{r}; t)$.

From Eq.~\eqref{eqn:dipole-conserving-Greens-function}, we find that
\begin{equation}
    Q \partial_x \partial_y G =
    \frac{Q}{D_0 t} \frac{xy}{r^2} \left[ f''\left( u \right) - \frac{1}{u} f'\left( u \right) \right]
    \, .
\end{equation}
Within perturbation theory, the dynamics of the $xy$ charge then follows immediately from dimensional analysis.
Each factor of $\rho_\text{q}(\vec{r}; t) = Q \partial_x \partial_y G$ introduces a factor of $t^{-1}$, while each spatial derivative gives a factor $t^{-1/4}$. Finally, the 2D integration measure introduces $t^{1/2}$. Any integral that depends on $\rho_\text{q}$ and derivatives thereof therefore scales as $t^{1/2-\#[\rho] - \#[\partial]/4} $. For the particular case of
$\mathcal{Q}_{xy}$, we find from (\ref{eq:isotropiccontinuumrelaxation}) that $\dot{\mathcal{Q}}_{xy} \sim t^{-3}$, leading to power-law relaxation or growth of $\mathcal{Q}_{xy} \sim -t^{-2} + \text{const.}$ depending on the sign of $D_2$. We emphasize that this integration constant is \emph{not zero} in general, as observed in Fig. \ref{fig:nonlinear-relxation}.

The predicted power-law scaling is observed in Fig.~\ref{fig:nonlinear-relxation}, where we simulate the nonlinear PDE defined by~\eqref{eqn:nonlinear-constitutive} subject to an initial charge density $\rho(\vec{r}; 0) = (q/\sigma^4) xy e^{-r^2/2\sigma^2}$, where $q$ is the charge enclosed in the quadrant $x,y>0$. The solution is obtained for times $D_0 t \ll L^4$ such that any boundary effects are negligible.
Note that there remains residual overlap in the long time limit; this is because the (irrelevant) non-linearity decays away sufficiently fast in time.
This should be contrasted with the lattice case discussed later (or indeed the case with boundaries), where the harmonic charges do appear to relax to zero at late times. 


\subsection{Relaxation with boundaries}
\label{sec:isotropic+boundaries}

On a domain of finite size, the boundary conditions also provide a mechanism by which the harmonic charges cease to be exactly conserved.
This mechanism is most crisply demonstrated by the simpler case of \emph{diffusive} hydrodynamics, in which only a \U1 charge is conserved.
The ubiquitous diffusion equation $\partial_t \rho - D \nabla^2 \rho = 0$ that follows from Fick's law actually conserves
\emph{both} charge \emph{and} dipole moment on the infinite plane.
This can be seen by writing the continuity equation for dipole moment density $x^\ell \rho$:
\begin{equation}
    \partial_t(x^\ell \rho) - D \partial^i ( x^\ell \partial_i \rho - \delta^{\ell}_i \rho) = 0
    \, .
    \label{eqn:charge-conserving-dipole-continuity}
\end{equation}
Boundary conditions that locally conserve charge demand that the normal derivative of charge density must vanish on the boundary, $\partial_{\vec{n}} \rho = 0$, where $\vec{n}$ is the unit vector normal to the boundary, and $\partial_\vec{n}$ is shorthand for $n^i\partial_i$.
However, it may be seen from~\eqref{eqn:charge-conserving-dipole-continuity} that a vanishing charge current is insufficient to preserve dipole moment at the boundary.

The situation is more subtle for the case of dipole-conserving hydrodynamics.
In the linear theory, the boundary conditions, $\nabla^2 \rho = 0$ and $\partial_{\vec{n}} \nabla^2 \rho = 0$, which are sufficient to conserve both charge and dipole moment, remarkably \emph{also} imply exact conservation of \emph{all} harmonic charges $\mathcal{Q}_f$ at the boundary.
Explicitly, if $f(\vec{r})$ is a harmonic function, $\laplacian f = 0$, then
\begin{align}
    \frac{\text{d}}{\text{d}t} \mathcal{Q}_f &= -D_0 \int \mathrm{d}^d \vec{r} \; f \nabla^4 \rho \notag \\
    &=
    - D_0 \int_{\partial V} \left[ f \partial_\vec{n} \laplacian \rho - (\laplacian\rho) \partial_\vec{n} f \right]
\end{align}
where in the second line we have integrated by parts.
This \emph{exact} conservation of the harmonic charges is illustrated in the bottom panel of Fig.~\ref{fig:cover_image}, where the charge $\mathcal{Q}_{xy}$, in addition to total charge and dipole moment, is shown to be constant in time.
The corresponding charge density at particular times during the evolution is depicted in the top panels, demonstrating that the conservation of $\mathcal{Q}_{xy}$ remains exact even once the charge density reaches, and hence interacts with, the system's boundaries.

If nonlinearities of the form~\eqref{eqn:nonlinear-constitutive} are included, then the two boundary conditions that are permitted by the fourth-order subdiffusion equation~\eqref{eq:introk4} become inadequate to describe the three exactly conserved quantities (in two dimensions): charge, and the two components of dipole moment.
The nonlinear corrections therefore necessitate including a term that is sixth order in spatial derivatives if we are to conserve all three of these quantities exactly.
Additionally, the nonlinear boundary conditions (like the bulk equations) no longer conserve higher-degree harmonic charges.

We emphasize that $\nabla^2 \rho = 0$ and $\partial_{\vec{n}} \nabla^2 \rho = 0$ are very special boundary conditions that preserve all of the relevant charges in the linear theory, but do not correspond to vanishing \emph{local} dipole flux $n_iJ^{ij}=0$: the two boundary conditions are related by a `solenoidal' current of the form $J^{ij} \to J^{ij} + \epsilon^{ik}\epsilon^{j\ell} \partial_k \partial_\ell \rho$, where $\epsilon^{ij}$ is the Levi--Cevita symbol.
Therefore, while the boundary conditions $\nabla^2 \rho = \partial_\vec{n} \nabla^2 \rho = 0$ imply that $\oint n_i J^{ij} = 0$, they give rise to a nonvanishing dipole flux locally of the form $n_iJ^{ij} = -n_i \epsilon^{ik}\epsilon^{j\ell} \partial_k \partial_\ell \rho$.
Hence, even within linear response, the harmonic charges can decay to zero at the longest time scales if vanishing dipole flux is enforced locally with the boundary conditions $n_i J^{ij}=0$.
The relaxation of the charges induced by the boundary conditions will be exponential in time at the longest timescales, $\sim e^{-cD_0 t L^{-4}}$ for some $c>0$, and commences once charge density has had sufficient time to subdiffuse to the boundary.
In principle, the possible importance of higher derivative terms at the boundary (as noted in the prior paragraph) suggests that there should be a `skin effect' and an emergent length scale near the boundary
where the charge is distributed in such a way as to conserve all components of dipole moment.
Unfortunately, we have not been able to see such a skin effect in our automaton numerics; the resolution thereof remains an outstanding problem for future work.


\section{Relaxation on the lattice}
\label{sec:relaxation-on-lattice}

We now turn to relaxation of harmonic function charges on a lattice. While the relaxation mechanism involving non-linear corrections to hydrodynamics, and its interplay with boundary effects, should also be operative on the lattice, we will show that in lattice systems there is an alternative mechanism, which
endows the harmonic charges with dynamics
as a bulk phenomenon. The key ingredient here is {\it lattice anisotropies}.  For generic harmonic charges, lattice anisotropies
permit dynamics
in the bulk, without need for either boundaries or non-linearities.\footnote{In principle there are alternative charges that are conserved in the bulk on average. For instance, for a subdiffusion equation $\partial_t \rho + D^{ijk\ell} \partial_i \partial_j \partial_k \partial_\ell \rho = 0$ the charges corresponding to functions $f$ solving $D^{ijk\ell} \partial_i \partial_j \partial_k \partial_\ell f = 0$ will be conserved in the bulk. However, these charges (unlike the harmonic charges) will necessarily be non-conserved in the presence of a boundary, and stochastic contributions to hydrodynamics (which must exist, by the fluctuation-dissipation theorem), even without any non-linear or higher derivative corrections.  We do not consider them further here.} A single charge and dipole conserving gate can relax these charges; however, they may be statistically conserved in the infinite plane (e.g., $\mathcal{Q}_{xy}$ on the square lattice), much as ordinary diffusion statistically conserves the dipole charge in infinite space. However, in the presence of boundaries and lattice anisotropies, these charges also become non-conserved, and relax to zero exponentially in time, with a relaxation timescale determined by the most relevant (i.e., fewest derivative) term exhibiting lattice anisotropy. This can lead to relaxation timescales longer than one would expect based on simple power counting, if the leading lattice anisotropies arise as dangerously irrelevant corrections to fracton hydrodynamics. 

\begin{figure}[t]
    \centering
    \includegraphics[width=\linewidth]{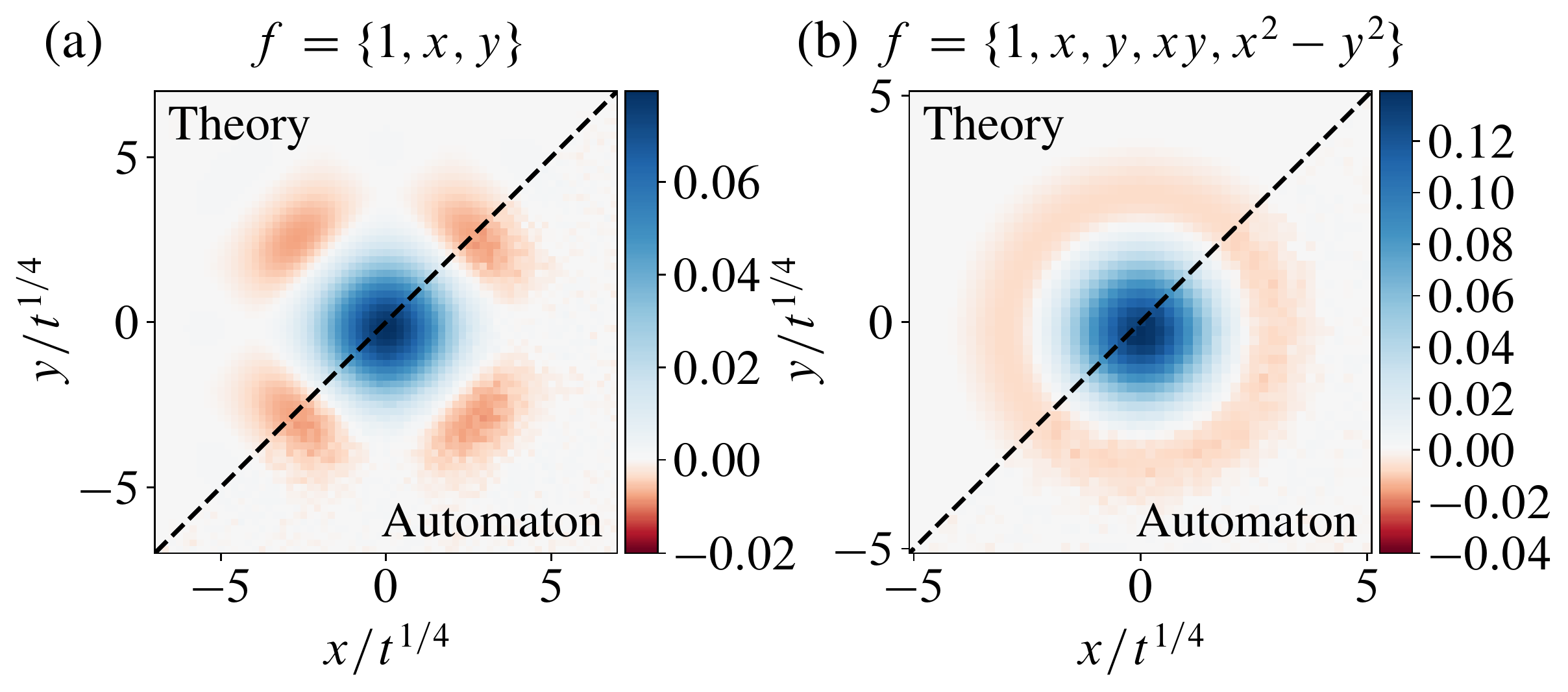}
    \caption{\textbf{Left:} autocorrelation function $\langle S^z (\mathbf{r}; t) S^z(\mathbf{0}; 0) \rangle$ for the dipole-conserving automaton circuit, which exhibits the same $D_4$ symmetry as the underlying lattice. The average is over initial conditions drawn from an infinite temperature ensemble, and $L_x=L_y=100$. \textbf{Right:} the same infinite temperature correlation function for the circuit that additionally conserves the traceless components of the quadruple tensor with $L_x=L_y=64$. In this case, isotropy is restored, and the correlation function exhibits an emergent rotational symmetry. The phenomenological constants $D^{xx}$ and $D^{xy}$ in Eq.~\eqref{eqn:Dijkl}, and $D_0$ in Eq.~\eqref{eqn:haah-scalar-J} are not used as fitting parameters, being extracted via alternative means, as described in Appendix~\ref{sec:extracting-D}. In both cases the automaton circuit employs gates of size $4 \times 4$, and the correlation function is plotted at a time $t=2^9$.}
    \label{fig:autocorrelator}
\end{figure}

We now elaborate on this high level summary. In the presence of an underlying lattice, the tensor structure permitted in the constitutive relation is nontrivial~\cite{fractonhydro}.
For instance, when the microscopic model lives on a square lattice,
the constitutive relation takes the form 
\begin{equation}
    J^{ij} = D^{ijk\ell} \partial_k \partial_\ell \rho
    \, ,
    \label{eqn:square-constitutive}
\end{equation}
where the $D_4$ symmetry of the square lattice enforces the following index structure:
\begin{equation}
    D^{ijk\ell} = (D^{xx}-3D^{xy}) \delta^{ijk\ell} + D^{xy} (\delta^{ij}\delta^{k\ell} +
     \delta^{ik}\delta^{j\ell} +
     \delta^{i\ell}\delta^{jk})
     \label{eqn:Dijkl}
\end{equation}
and $\delta^{ijkl} = \delta_{ij}\delta_{ik}\delta_{i\ell}$ sets all indices equal to one another.
With periodic boundary conditions, a modulation of the density with wave vector $\vec{k}$, $\rho(\vec{r}) = \rho_0 e^{i\vec{k}\cdot \vec{r}}$, therefore decays with time asymptotically as $e^{-\Gamma(k_x, k_y)t}$, where
\begin{equation}
    \Gamma(k_x, k_y) = D^{xx}(k_x^4 + k_y^4) + 6 D^{xy} k_x^2 k_y^2
    \, .
    \label{eqn:square-lattice-D}
\end{equation}
This prediction for the anisotropic decay rate is confirmed by the infinite temperature correlation function $C_z(\vec{r}; t)$ in Fig.~\ref{fig:autocorrelator} (a).
The ratio $D^{xy}/D^{xx}$ can be inferred from the correlation function, or, to a higher degree of accuracy, by performing a scaling collapse of the imbalance $\mathcal{I}(t) \sim e^{-\Gamma(k_x, k_y)t}$ for various wave vectors $\vec{k}$. The latter approach gives $D^{xy}/D^{xx} = 1.1(3)$, consistent with Ref.~\cite{IaconisAnyDimension2021}.
Using this ratio, we observe very good agreement between the hydrodynamic prediction~\eqref{eqn:hydro-correlator} and the automaton circuit with no adjustable parameters.

The boundary conditions that are relevant to the automaton dynamics of circuits with open boundaries are zero local charge flux and dipole flux across the boundary, $n_i\partial_j J^{ij}=0$ and $n_iJ^{ij}=0$, respectively.
As shown in Appendix~\ref{sec:n-pole-continuity}, the time dependence of the complex harmonic charge $\mathcal{Q}_n = \int\mathrm{d}^2\mathbf{r} \; (x+iy)^n \rho$ in the presence of such zero flux boundary conditions is given by
\begin{equation}
    \frac{\mathrm{d}}{\mathrm{d}t} \mathcal{Q}_n = - n(n-1) \int \mathrm{d}^2\mathbf{r} \;  (J^{xx}-J^{yy}+2iJ^{xy}) z^{n-2}
    \, ,
    \label{eqn:complex-charge-eom-dipole}
\end{equation}
with $z=x+iy$. The equation for the $xy$ charge, Eq.~\eqref{eq:isotropiccontinuumrelaxation}, is recovered by taking the imaginary part of the above equation for $n=2$.
At the linear level, the constitutive relation~\eqref{eqn:square-constitutive} gives $J^{xy} = 2D^{xy} \partial_x \partial_y \rho$.
Performing the integral over space, it follows that
\begin{equation}
    \frac{\mathrm{d}}{\mathrm{d}t} \mathcal{Q}_{xy} = -4D^{xy} \sum_{\substack{\vec{r} \in \\ \text{corners}}} (-1)^c \rho(\vec{r}; t)
    \, ,
    \label{eqn:Qxy-eom}
\end{equation}
where, for the domain $-1<x<1$, $-1<y<1$, we have $c=\sign x\sign y$.
The zero flux boundary conditions do not prevent a nonzero charge density at the corners, and so $\mathcal{Q}_{xy}$ will inevitably decay at sufficiently long times.
For the $n=3$ $(n \geq 4)$ charges, in the analogue of~\eqref{eqn:Qxy-eom}, the corner contribution is replaced by a boundary (bulk) term, but the conclusion remains unaltered: The charges $\mathcal{Q}_n$ will, at sufficiently long times, decay exponentially, with a rate given by the slowest hydrodynamic mode compatible with the boundary conditions with which the initial conditions have nonzero overlap.
When the equation of motion includes a bulk contribution, the harmonic charges can exhibit dynamics before the unavoidable relaxation that occurs on the longest time scales equal to the time taken to subdiffuse to the boundary.

\begin{figure}[t]
    \centering
    \includegraphics[width=\linewidth]{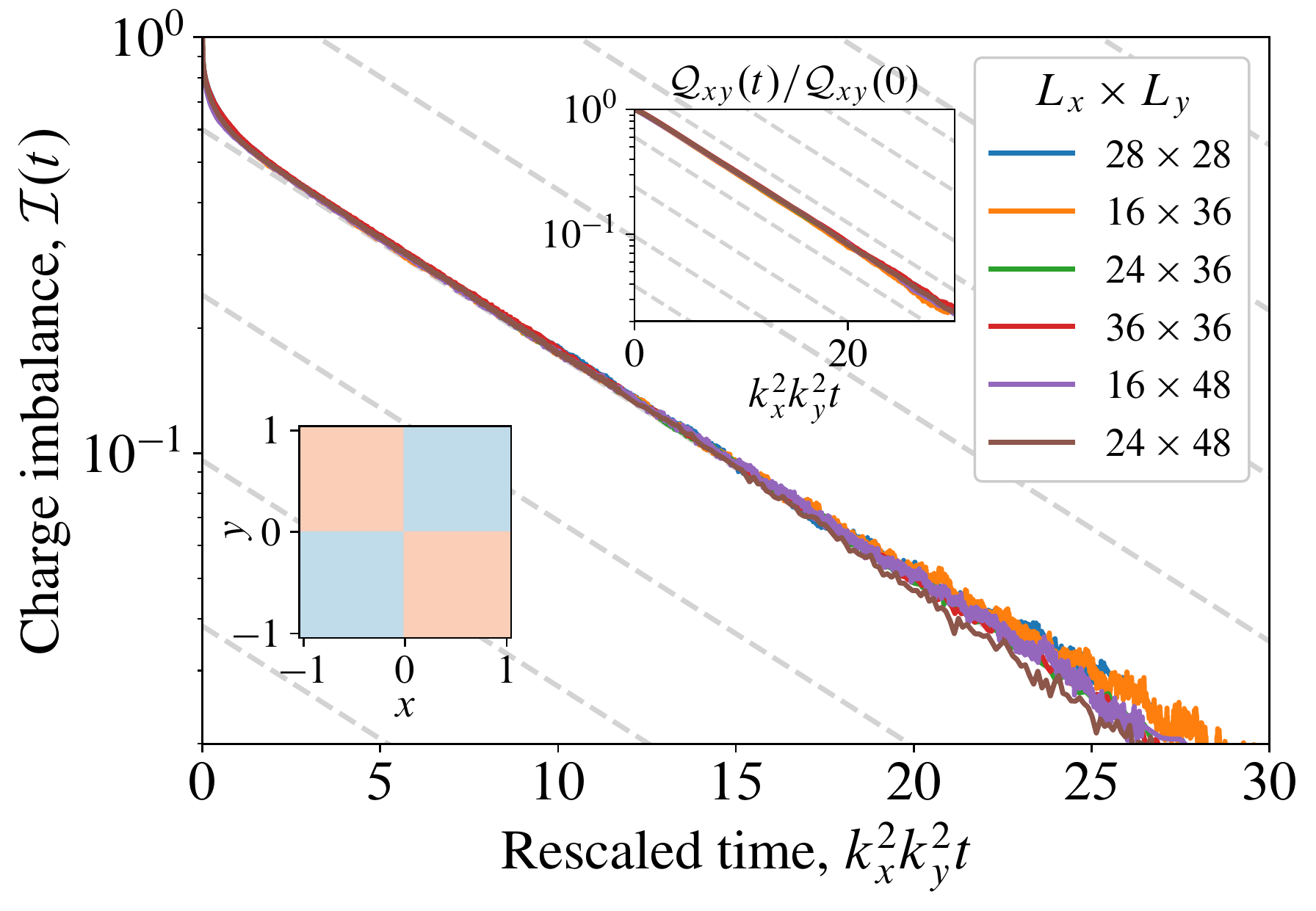}
    \caption{Relaxation of the charge imbalance $\mathcal{I}(t)$ in the dipole conserving automaton circuit with open boundary conditions, starting from the quadrupole initial state (shown in the bottom left inset). This initial state has large overlap with the $xy$ charge $\mathcal{Q}_{xy}$. The near perfect scaling collapse identifies asymptotic exponential decay, $e^{-t/\tau_{xy}}$, with $\tau_{xy} \propto k_x^{-2} k_y^{-2}$, and $k_i = 2\pi/L_i$. The decay of the $xy$ charge $\mathcal{Q}_{xy}$ is shown in the top right inset, which also exhibits exponential decay with the same time scale $\tau_{xy}$. The automaton circuit uses gates of size $4 \times 4$, and the curves are averaged over at least $2^{14}$ initial states.}
    \label{fig:k4-collapse-OBC}
\end{figure}

These predictions are consistent with what is observed in the automaton circuits.
As shown in Fig.~\ref{fig:k4-collapse-OBC}, when open boundary conditions are imposed on the circuit, the imbalance decays to zero asymptotically as $\mathcal{I}(t) \sim e^{-t/\tau}$, where\footnote{The discrepancy between this expression and Eq.~\eqref{eqn:square-lattice-D} is explained by the boundary conditions. The relaxation of a mode with wavelength $\lambda \ll L$ will relax according to~\eqref{eqn:square-lattice-D}, but this relationship ceases to hold for wavelengths $\lambda \sim L$.} $\tau \propto L_x^2 L_y^2$.
This timescale is indeed shared by the decay of the $xy$ charge $\mathcal{Q}_{xy}\sim e^{-t/\tau}$, as shown in the inset of Fig.~\ref{fig:k4-collapse-OBC}.

In the example discussed above, the relaxation time for the $xy$ charge is what one might have guessed based on simple power counting, noting that boundaries are important for relaxation and dipole conserving hydrodynamics has $k^4$ subdiffusion. We now turn to a richer example, where the fracton hydrodynamics is isotropic at leading order, and the lattice anisotropies are {\it dangerously irrelevant} perturbations to the leading order isotropic fracton hydrodynamics.


\subsection{Relaxation from dangerously irrelevant lattice anisotropies}
\label{sec:xy-conserving}

We now consider a three-dimensional fluid inspired by the $\text{U}(1)$ Haah code~\cite{BulmashBarkeshli2018,Gromov2019,Gromov2020duality,WesleiChamon2021,Hart2021TypeII} in which the following functions $f(x,y,z)$ must be exactly conserved:
\begin{equation}
    f=\lbrace 1,\, x,\, y,\, xy,\, x^2-y^2\rbrace
    \, ,
    \label{eqn:conserved-fs}
\end{equation}
corresponding to charge, $x$ and $y$ dipole moment, and the traceless components of the quadrupole tensor in the $xy$-plane. 
(Note that we are rotating coordinates from the usual choice in the microscopic Haah code).
Within the $xy$-plane, an isolated charge is immobile, while an isolated dipole can move only along its dipole moment~\cite{Gromov2020duality}.
In general, one expects that such a theory should admit a hydrodynamic description of the form
\begin{equation}
    \partial_t \rho + \sum_\mu D_\mu J_\mu = 0
    \, ,
    \label{eqn:generalised-derivative-continuity}
\end{equation}
where the $D_\mu$ are generalised derivative operators \cite{Gromov2019}, and the $J_\mu$ are the currents to which they couple.
Note that the index $\mu$ is not necessarily related to any spatial indices.
In the absence of boundaries, the generalised continuity equation~\eqref{eqn:generalised-derivative-continuity} implies that 
\begin{equation}
    \frac{\mathrm{d}}{\mathrm{d}t} \int \mathrm{d}^d \vec{r}\;
    P(\vec{r}) \rho(\vec{r}; t) = 0
    \, ,
\end{equation}
if the functions $P(\vec{r})$ satisfy $D^\dagger_\mu P(\vec{r}) = 0$ for all $\mu$.
The operators $D^\dagger_\mu$ are related to the $D_\mu$ via integration by parts, $\int \mathrm{d}^d\vec{r}\; f D_\mu g = \int\mathrm{d}^d\vec{r} \; g D_\mu^\dagger f$, up to boundary terms.
If the system conserves the multipole moments in~\eqref{eqn:conserved-fs}, then to lowest order in spatial derivatives, the most general set of $D_\mu$ that one can write down is
\begin{equation}
    \partial_t \rho + \partial_z J_z + \nabla_\perp^2 J + \partial_A \partial_B \partial_C J_{ABC} = 0
    \, ,
    \label{eqn:Haah-fluid}
\end{equation}
where $\nabla^2_\perp = \partial_x^2+\partial_y^2$ and the indices $A,B,C$ run over $x$ and $y$ only.
A more formal and extended discussion of this fact is a topic for future work~\cite{marvin}.
Focusing on the dynamics in the $xy$ plane, we find that the leading-order hydrodynamic theory has a \emph{scalar} current
\begin{equation}
    J = D_0 \nabla^2_\perp \rho
    \, ,
    \label{eqn:haah-scalar-J}
\end{equation}
which leads to fourth-order subdiffusion as if only dipoles were conserved. Crucially, however, at the leading derivative level, this theory \emph{exactly} conserves all harmonic charges, even at the nonlinear level; as a consequence, the sixth order corrections to the hydrodynamic equations of motion
\begin{equation}
    J_{ABC} \sim -D^\prime \partial_A \partial_B \partial_C \rho
\end{equation}
represent dangerously irrelevant operators even within linear response.  We numerically demonstrate that it is these sixth-order subdiffusive corrections that relax the non-conserved harmonic charges in the ``Haah code fluid''.  This represents a peculiar realization of ``UV/IR mixing'' in type-II fracton matter \cite{BulmashBarkeshli2018, Gromov2019, Hart2021TypeII, Schmit}, in which microscopic length scales explicitly show up in the hydrodynamic theory.

We now elaborate on this high level summary, focusing on the behavior of the Haah code fluid in the $xy$-plane.
We first consider the behaviour of the infinite temperature correlation function
\begin{align}
        C_z(\vec{r}; t) &\simeq \frac{1}{3}S(S+1) \int \frac{\mathrm{d}^2 \vec{k}}{(2\pi)^2} e^{i\vec{k}\cdot\vec{r}} e^{-D_0 k^4 t} \notag \\
        &= \frac{S(S+1)}{3\sqrt{D_0t}}  f\left( u \right)
        \, ,
        \label{eqn:Cz-haah}
\end{align}
where $u=r(D_0t)^{-1/4}$, and the function $f(u)$ is defined in Eq.~\eqref{eqn:dipole-conserving-Greens-function}.
In Fig.~\ref{fig:autocorrelator}~(b), this exact expression is compared with the results from a 2D automaton circuit that utilises local gates $\hat{U}$ that exactly conserve the multipole moments in~\eqref{eqn:conserved-fs} relevant to subdiffusion in the $xy$-plane: charge, dipole moment, and the traceless components of the quadrupole tensor.
We observe essentially perfect agreement between the numerics and Eq.~\eqref{eqn:Cz-haah}.
In principle, the plot contains just one adjustable parameter, the phenomenological constant $D_0 > 0$, which relates charge and current in the automaton circuit. However, the parameter $D_0$ can be extracted via independent means by performing a scaling collapse of the imbalance $\mathcal{I}(t) \sim e^{-\Gamma(\vec{k})t}$ for various wave vectors $\vec{k}$ (see Appendix~\ref{sec:extracting-D}).
This leaves no free parameters in Fig.~\ref{fig:autocorrelator}.

Having established that, at the leading derivative level, the $xy$ and $x^2-y^2$ conserving circuit exhibits \emph{isotropic}
subdiffusion, we now examine the fate of the harmonic charges $\mathcal{Q}_f$.
The boundary conditions appropriate for zero local dipole, charge, and quadrupole flux are $J=\partial_\vec{n} J = 0$.
Explicitly, the local harmonic charge density evolves according to
\begin{equation}
    \partial_t (f\rho) + \partial^i (f\partial_i J - J \partial_i f) = -J\partial_i\partial^i f
    \, ,    
\end{equation}
where the source term vanishes for harmonic functions. When integrated over the appropriate domain, the boundary term vanishes for the aforementioned zero flux boundary conditions.

The additional conservation laws enforce that the rank-four 
tensor appearing in the hydrodynamic description must transform trivially under the point group symmetry of the lattice.
However, as noted at the beginning of this section, there also exist terms higher order in spatial derivatives consistent with the conservation laws defined by Eq.~\eqref{eqn:conserved-fs}.
These higher-order corrections in general will transform nontrivially under the point group.
For instance, on the square lattice, the permitted index structure in the constitutive relation is
\begin{equation}
    J^{ijk} = D^{ijk\ell m n} \partial_\ell \partial_m \partial_n \rho
    \, ,
\end{equation}
where the rank-six tensor $D^{ijk\ell mn}$ contains two independent parameters
\begin{equation}
    D^{ijk\ell mn} =
    \begin{cases}
        D_1 & \text{6 matching indices} \\
        D_2 & \text{4 matching indices}
    \end{cases}
    \, .
\end{equation}
The anisotropic sixth-order term therefore has the ability to relax the harmonic charges at the longest time scales, just as lattice anisotropies combined with boundaries were responsible for the relaxation of $n>1$ harmonic charges in Sec.~\ref{sec:relaxation-on-lattice}.

As shown in Appendix~\ref{sec:n-pole-continuity}, the explicit equation of motion for the complex harmonic charges $\mathcal{Q}_n$ in the presence of boundaries is
\begin{multline}
    \frac{\mathrm{d}}{\mathrm{d}t} \mathcal{Q}_n
    =
    n(n-1)(n-2) \times \\
    \int \mathrm{d}^2 \mathbf{r} \; \left[ J^{xxx} + 3i J^{xxy} -3 J^{xyy} -iJ^{yyy} \right] z^{n-3}
    \, ,
\end{multline}
with $z=x+iy$.
\begin{figure}[t]
    \centering
    \includegraphics[width=\linewidth]{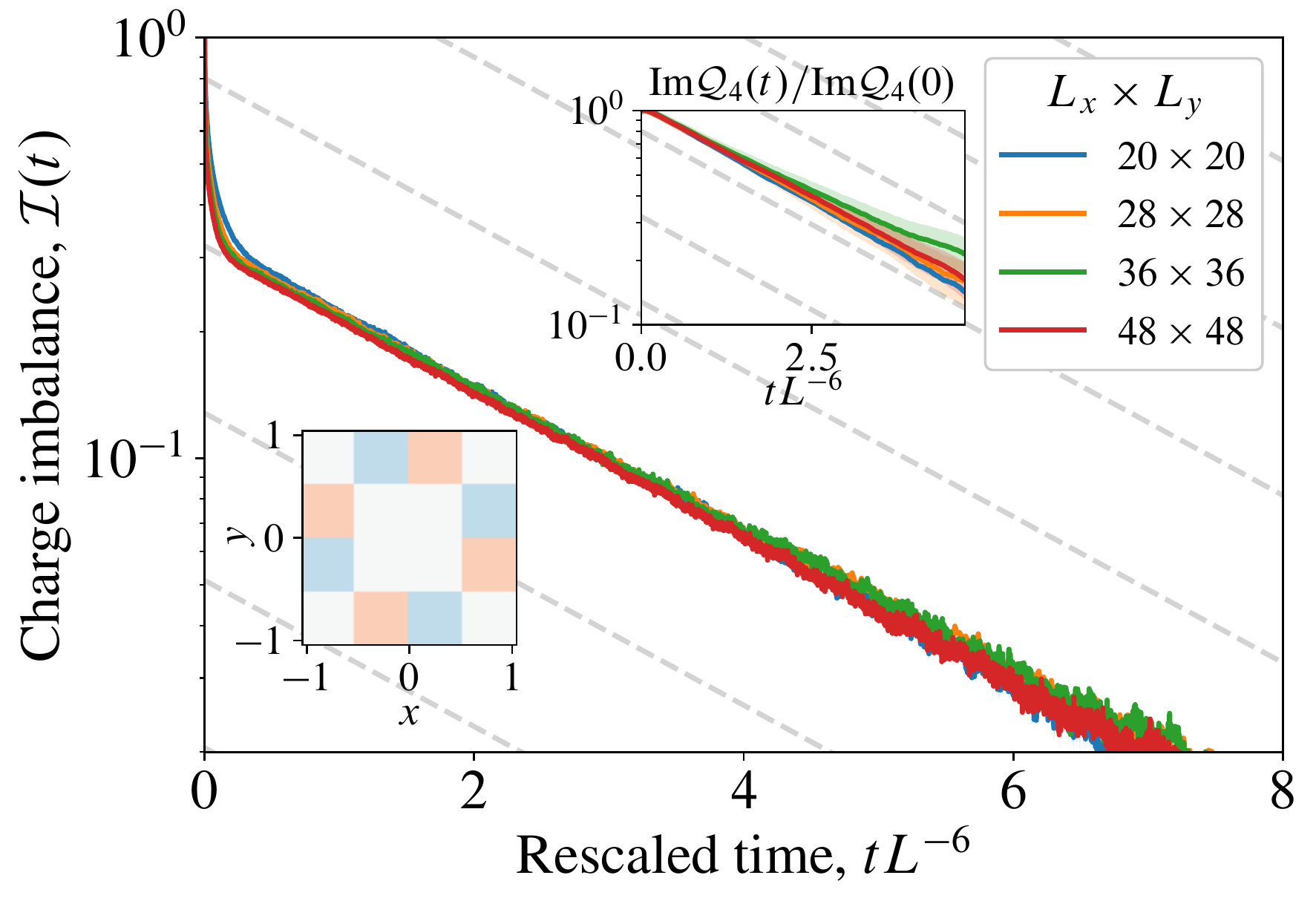}
    \caption{Relaxation of the charge imbalance $\mathcal{I}(t)$ in the traceless quadrupole conserving automaton circuit, starting from a four-pole initial state (shown in the lower left inset). The scaling collapse occurs when time is rescaled by $t\to L^{-6} t$, consistent with the arguments presented in the main text that the anisotropic sixth order (quadrupole-conserving) terms in Eq.~\eqref{eqn:Haah-fluid} are responsible for the relaxation of harmonic charges (of degree $n > 2$). The decay of the four-pole charge $\Im \mathcal{Q}_4$ is plotted in the upper right inset, and is consistent with the same decay rate as the imbalance. The curves are averaged over at least $2^{14}$ initial states.}
    \label{fig:sixth-order-relaxation}
\end{figure}
As explained above, absent the higher-rank tensor $J^{ijk}$, the right hand side vanishes.
When it is included, the right-hand side of the above equation is generally nonzero for harmonic charges with degree higher than those which are \emph{exactly} conserved by the dynamics (here $n>2$),  just as in Eq.~\eqref{eqn:complex-charge-eom-dipole}.
At the longest time scales, the harmonic charges should decay exponentially $\sim e^{-\Gamma(L_x, L_y) t}$ with the relaxation rate $\Gamma$ set by the slowest hydrodynamic mode. In the presence of the higher derivative corrections, we anticipate that the corresponding gap should scale as $k^6$. This prediction is tested in the 2D $xy$ and $x^2-y^2$ conserving automaton circuit in Fig.~\ref{fig:sixth-order-relaxation}. We initialise the circuit in a ``four-pole'' initial state (see Table~\ref{tab:polynomials}) that has vanishing charge, dipole, $xy$ and $x^2-y^2$ moments, but nonzero four-pole charge.
As expected in light of the above arguments, the imbalance and the four-pole charge $\Im \mathcal{Q}_4$ both decay exponentially in time with an identical decay rate consistent with the scaling $\Gamma \sim L^{-6}$.


\section{Conclusion and outlook}
\label{sec:outlook}

We have pointed out that fracton hydrodynamics can have a large number of hidden `quasiconserved charges'. In our examples, these extra quasiconserved charges are associated to harmonic functions. Their conservation is not put in explicitly, but emerges from the constrained form of the hydrodynamic equations. If working in infinite isotropic continuum space, some fraction of these charges will not decay, even in the infinite time limit, although non-linear corrections to hydrodynamics do cause some algebraic in time decay or growth. In the presence of boundaries, these charges can be relaxed at the level of linear hydrodynamics depending on the boundary conditions.
When moving to a \emph{lattice} there is a further wrinkle coming from lattice anisotropies. Lattice anisotropies render most of the charges non-conserved even in the bulk, but finitely many harmonic charges remain conserved in an infinite lattice. On a lattice \emph{with boundaries} however, all the harmonic charges are able to relax. The relaxation time is set by the leading lattice anisotropy. In settings where the leading anisotropic term carries more derivatives than the leading term in fracton hydrodynamics, the relaxation comes from dangerously irrelevant operators, and is slower than naive power counting would predict. We have illustrated these results by both numerical solution of PDEs and automaton Monte Carlo simulations, including on an explicit lattice model motivated by the U(1) Haah code. 

The phenomenon we have discovered might evoke analogies to long-time tails in conventional hydrodynamics \cite{PhysRevA.27.3216,Kovtun:2003vj}, wherein hydrodynamic fluctuations lead to the anomalously slow decay of certain degrees of freedom (e.g., non-conserved operators have power-law, rather than exponential, decay with time, similar to the decay in Fig.~\ref{fig:nonlinear-relxation}).  However, the effect is not the same: in our case, one does not need to consider ``loop corrections" within nonlinear fluctuating hydrodynamics to see the anomalous power-law decays; they are already present within the classical PDEs.  In contrast, our accidental conservation laws primarily stem from the curious higher-derivative nature of the currents. 

Could these quasiconservation laws be seen experimentally? Atomic, molecular, and optical (AMO) platforms appear the most promising candidates for such explorations in the near future, and identification of a specific platform suitable for realization would be an interesting topic for future work. We note also that the phenomena discussed herein could also be relevant for quantum magnetism with large effective spins. For example, phenomena analogous to the discussion in Sec.~\ref{sec:xy-conserving} arise in spin-$S$ lattice models endowed with a local Hamiltonian that conserves the moments $\hat{\mathcal{Q}}_f = \sum_{i} f(\vec{r}) \hat{S}_i^z$ defined by the functions $f(\vec{r})$ in~\eqref{eqn:conserved-fs}. 
Some of the simplest local stencils that one can write down that respect the conservation laws~\eqref{eqn:conserved-fs}, \emph{also} conserve an infinite hierarchy of harmonic charges (on a finite lattice, the number of such charges scales with the number of boundary sites, $\propto L^{d-1}$).
This is most simply demonstrated by the following $d=2$ local Hamiltonians, $\hat{H}^{(3)}$ and $\hat{H}^{(5)}$, which are defined on the square lattice, with spin-$S$ degrees of freedom on the sites, and are constructed from $3\times 3$ and $5 \times 5$ stencils, respectively:
\begin{subequations}\label{eqn:conserved-Q-from-Hamiltonian}
    \begin{align}
        \hat{H}^{(3)} &= J\sum_\vec{r} \left( \prod_{\boldsymbol{\delta}\in \{\pm \vec{e}_x, \pm \vec{e}_y\}} \hat{S}_{\vec{r}}^+ \hat{S}_{\vec{r}+\boldsymbol{\delta}}^- + \text{H.c.} \right)
        \, , \\
        \hat{H}^{(5)} &= J\sum_\vec{r} \left( \prod_{\boldsymbol{\delta}\in \{\pm \vec{e}_x, \pm \vec{e}_y\}} \hat{S}_{\vec{r}+\boldsymbol{\delta}}^+ \hat{S}_{\vec{r}+2\boldsymbol{\delta}}^- + \text{H.c.} \right)
        \, .
    \end{align}
\end{subequations}
Note that $S_\vec{r}^\pm$ appears to the fourth power in $\hat{H}^{(3)}$, and so only acts nontrivially on spins with $S \geq 2$.
As shown in Appendix~\ref{sec:local-hamiltonians}, for the Hamiltonians in~\eqref{eqn:conserved-Q-from-Hamiltonian}, $[\hat{H}^{(n)}, \hat{\mathcal{Q}}_f] = 0$ for any discrete function $f(\vec{r})$ that satisfies the discretised Laplace equation, $\sum_{\vec{r}'} \Delta_{\vec{r}\vec{r}'} f(\vec{r}') = 0$, where $\Delta_{\vec{r}\vec{r}'}$ is defined by the stencil that appears in~\eqref{eqn:conserved-Q-from-Hamiltonian}.
For instance, $\hat{H}^{(3)}$ corresponds to the canonical ``five-point stencil'' finite difference approximation to the Laplacian
\begin{equation}
    (\Delta f)_{x,y} = f_{x+1, y} + f_{x-1, y} + f_{x, y+1}+ f_{x, y-1} - 4f_{x, y}
    \, .
\end{equation}
Since we are primarily concerned with the \emph{relaxation} of the harmonic charges, while they are conserved exactly by the Hamiltonians in~\eqref{eqn:conserved-Q-from-Hamiltonian}, we will not discuss them further, except to note that the identification of materials that might plausibly realize such effective Hamiltonians would be an interesting topic for future work. 

Another question that remains open is how exactly the quasiconserved charges discussed herein interact with the shattering of Hilbert space discussed in Refs.~\cite{KhemaniShattering, SalaFragmentation}? Do some of the states where the harmonic charges are anomalously large end up lying in Krylov subsectors of finite size, and therefore fail to thermalize entirely? As a related point, one could ask: What are the simplest terms we can write down that couple harmonic charge sectors within a Krylov subsector? What about within a symmetry sector?  Answers to these detailed questions are interesting open questions. Regardless of their answers, our work reveals that fracton hydrodynamics remains very far from being fully understood, and may have yet more surprises in store.


\section*{Acknowledgements}
This material is based in part (O.H., R.N.) upon work supported by the
Air Force Office of Scientific Research under award number FA9550-20-1-0222. R.N.~also acknowledges the support of the Alfred P.~Sloan Foundation through a Sloan Research Fellowship.
A.L.~was supported in part by the Alfred P.~Sloan Foundation through Grant FG-2020-13795, the National Science Foundation under CAREER Grant DMR-2145544, and through the Gordon and Betty Moore Foundation's EPiQS Initiative via Grant GBMF10279.\\

\emph{Note added.}---Recently, we learned of an independent numerical study of the $xy$ and $x^2-y^2$ conserving model in an upcoming paper~\cite{sala2021dynamics}.

\appendix
\renewcommand{\thesubsection}{\thesection.\arabic{subsection}}
\renewcommand{\thesubsubsection}{\thesubsection.\arabic{subsubsection}}


\begin{figure*}[t]
    \centering
    \includegraphics[width=0.45\linewidth]{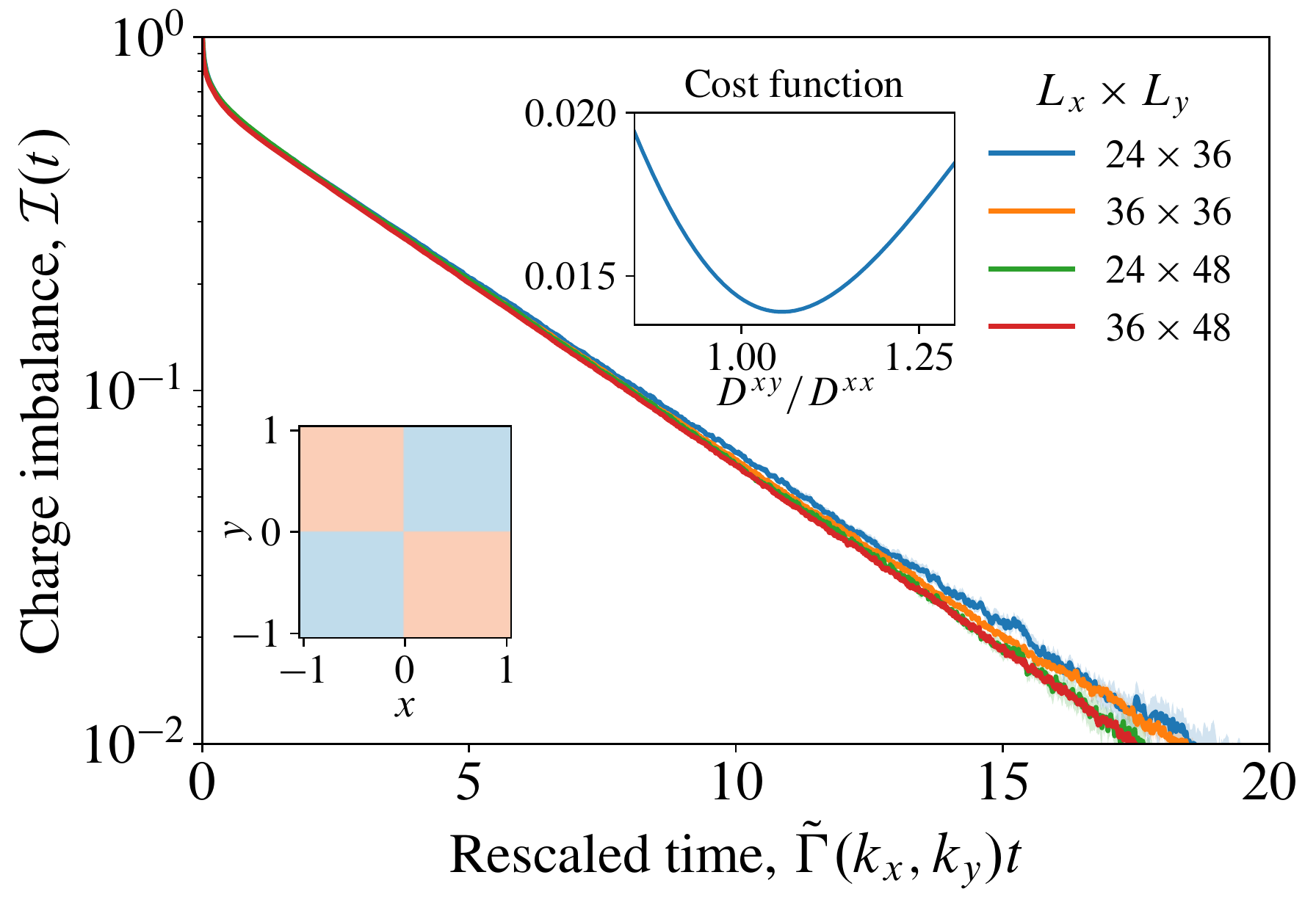}
    \hspace{1cm}
    \includegraphics[width=0.45\linewidth]{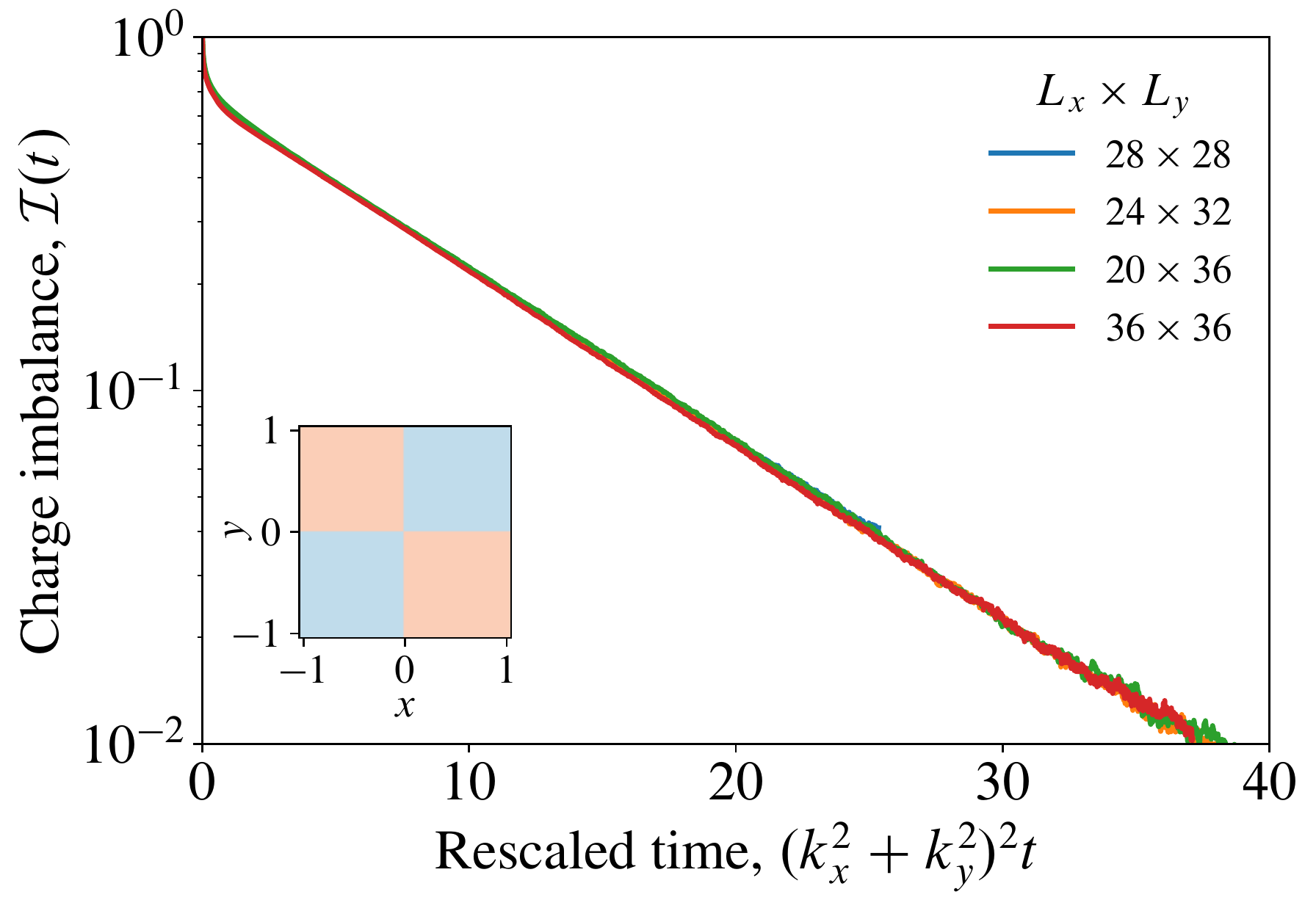}
    \caption{\textbf{Left:} Scaling collapse of the imbalance, $\mathcal{I}(t)$, in the dipole conserving automaton circuit for a quadrupole initial state (as shown in the inset). The optimal scaling collapse occurs for $\tilde{\Gamma}(k_x, k_y) = k_x^4 + k_y^4 + 6 \alpha k_x^2 k_y^2$, with $\alpha \equiv D^{xy}/D^{xx} \simeq 1.1$, and $k_i = 2\pi/L_i$. The resulting exponential decay is best fit by $D^{xx} \simeq 0.25$. \textbf{Right:} Scaling collapse of the imbalance for the $xy$ and $x^2-y^2$ conserving automaton circuit. The data exhibits a perfect scaling collapse when time is rescaled by the isotropic factor $(k_x^2 + k_y^2)^2$. The best fit of the corresponding exponential decay gives $D_0 \simeq 0.114$. Both circuits make use of \emph{periodic} boundary conditions, in contrast to earlier figures.}
    \label{fig:PBC-decay}
\end{figure*}

\section{Extracting phenomenological constants}
\label{sec:extracting-D}

In order to determine the phenomenological parameters entering the hydrodynamic equations accurately, we look at the relaxation of density modulations with various wave vectors $\vec{k}$.


\subsection{Dipole conserving circuits}

In dipole conserving circuits subjected to periodic boundary conditions, the relaxation of a density modulation of wave vector $\vec{k}$ is $\sim e^{-\Gamma(\vec{k})t}$, where $\Gamma(\vec{k}) = D^{ijk\ell} k_i k_j k_k k_\ell$.
As stated in the main text, on the square lattice, the permitted index structure in $D^{ijk\ell}$ is~\cite{IaconisAnyDimension2021}
\begin{equation}
    D^{ijk\ell} = (D^{xx}-3D^{xy}) \delta^{ijk\ell} + D^{xy} (\delta^{ij}\delta^{k\ell} +
     \delta^{ik}\delta^{j\ell} +
     \delta^{i\ell}\delta^{jk})
     \, .
\end{equation}
To extract $D^{xy}$ and $D^{xx}$, we perform a scaling collapse of the imbalance $\mathcal{I}(t)$ starting from quadrupole initial states with wave lengths $\lambda \gg 1$, and with various $k_x/k_y$, as performed in Ref.~\cite{IaconisAnyDimension2021} for a layered circuit architecture.
The optimal collapse is found by varying the ratio $D^{xy}/D^{xx}$, and determining the value for which the spread of the curves is minimised.
More precisely, using linear interpolation we sample $-\log \mathcal{I}$ at $M$ equally spaced points and, at each point, compute the standard deviation of the times relative to the average. The average over all $M$ points can then be used as a non-negative cost function, whose minimum occurs at the optimal value of $D^{xy}/D^{xx}$. The curvature of the parabolic minimum can also be used to extract the associated error in the optimal value.
The cost function and the associated scaling collapse are shown in Fig.~\ref{fig:PBC-decay}.
We find that the optimal collapse occurs for $D^{xy}/D^{xx} = 1.1(3)$. Note that the origin of the error is predominantly systematic in nature; the residual spread of the data at the optimal value of $D^{xy}/D^{xx}$ is ostensibly due to corrections higher order in $k$, which could in principle be removed by accessing larger system sizes. Having established the relative magnitude of $D^{xy}$ and $D^{xx}$, their overall scale can be determined by finding the gradient of the collapsed data.


\subsection{Traceless quadrupole conserving circuits}

The procedure for extracting $D_0$ is trivial, since the relaxation rate $\Gamma(\vec{k}) \propto k^4$ is isotropic. Hence, no additional step to collapse the data is necessary. As shown in Fig.~\ref{fig:PBC-decay}, the data collapse is essentially perfect using a fourth-order isotropic $\Gamma(\vec{k})$, even for the modest system sizes considered.
The value of $D_0$ can then be obtained by fitting the exponential tail, resulting in $D_0 \simeq 0.114$.

\begin{widetext}


\section{$n$-pole continuity equation}
\label{sec:n-pole-continuity}


\subsection{Dipole conserving hydrodynamics}

In this appendix we derive the continuity equation for a generic $n$-pole moment of the charge density for dipole conserving fracton hydrodynamics.
Our starting point is the continuity equation for $\rho$ (recall that the $i_n$ are indices, not exponents)
\begin{equation}
    \partial_t (x^{i_1} x^{i_2} \cdots x^{i_n} \rho) + x^{i_1} x^{i_2} \cdots x^{i_n} \partial_a \partial_b J^{ab} = 0
    \, ,
\end{equation}
where $a,b \in \{ 1,\ldots, d \}$.
Moving the product $x^{i_1} \cdots x^{i_n}$ under the spatial derivative, we arrive at
\begin{equation}
    \partial_t (x^{i_1} \cdots x^{i_n} \rho) + \partial_a \left( x^{i_1} \cdots x^{i_n} \partial_b J^{ab} - n x^{(i_1} \cdots x^{i_{n-1}} J^{i_n) a} \right) + n(n-1) x^{(i_1} \cdots x^{i_{n-2}} J^{i_{n-1}i_n)} = 0
    \, .
    \label{eqn:n-pole_dipole}
\end{equation}
The notation $T^{(ab)}$ indicates symmetrisation over the indices within the brackets, e.g., $T^{(ab)}=\frac12(T^{ab}+T^{ba})$.
For $n\geq 2$, i.e., quadrupoles and higher-degree polynomials, the continuity equation~\eqref{eqn:n-pole_dipole} contains a source term.
We now specialise to $d=2$ for the remainder of the derivation.
The equation of motion for the harmonic charges follows directly from~\eqref{eqn:n-pole_dipole}. It will be convenient to work with a complex representation of the harmonic charges $\mathcal{Q}_n \equiv \int \mathrm{d}^2 \vec{r} \; (x+iy)^n \rho$. We therefore require
\begin{equation}
    \frac{\mathrm{d}}{\mathrm{d}t} \mathcal{Q}_n = \int \mathrm{d}^2 \vec{r} \; \partial_t \left[ (x+iy)^n \rho(\vec{r}; t) \right] 
    \, .
\end{equation}
Expanding the harmonic polynomials $(x+iy)^n = \sum_{k=0}^n \binom{n}{k}i^k y^k x^{n-k}$, each term in the expansion can be evaluated using~\eqref{eqn:n-pole_dipole}. If zero charge and dipole flux boundary conditions are imposed, i.e., $n_a\partial_b J^{ab}=0$ and $n_aJ^{ab}=0$ on the boundary, then the surface terms (which are proportional to the charge and dipole flux with appropriate factors of $x^i$) vanish. The only nontrivial contribution therefore comes from the source term $\propto x^{(i_1} \cdots x^{i_{n-2}} J^{i_{n-1}i_n)}$.
This bulk contribution evaluates to
\begin{equation}
    \partial_t \left[ (x+iy)^n \rho \right] \supset
    -\sum_{k=0}^n \binom{n}{k}i^k y^{k}x^{n-k} \left[ J^{xx}(n-k)(n-k-1)x^{-2} + 2J^{xy}k(n-k)x^{-1}y^{-1} + J^{yy}k(k-1)y^{-2}  \right]
    \, .
\end{equation}
Now, by relabelling $k\to k+1$ ($k+2$) in the second (third) term in the square brackets, the expression can be resummed, giving
\begin{equation}
    \partial_t \left[ (x+iy)^n \rho \right] \supset -n(n-1) (x+iy)^{n-2} (J^{xx} - J^{yy} + 2iJ^{xy})
    \, .
\end{equation}
Integrating over space gives Eq.~\eqref{eqn:complex-charge-eom-dipole} presented in the main text.


\subsection{Traceless quadrupole conserving hydrodynamics}

In a similar spirit to the previous subsection, we derive the continuity equation for a generic $n$-pole moment of the charge density in the circuit that additionally conserves $xy$ and $x^2 - y^2$ moments of the charge density.
As before, we begin with the most general continuity equation for $\rho$
\begin{equation}
    \partial_t (x^{i_1} x^{i_2} \cdots x^{i_n} \rho) + x^{i_1} x^{i_2} \cdots x^{i_n} \left( \laplacian J + \partial_a \partial_b \partial_c J^{abc} \right) = 0
    \, .
    \label{eqn:generic-n-pole-continuity}
\end{equation}
As discussed in the main text, it is important to include \emph{both} the isotropic part \emph{and} the rank-three tensor $J^{abc}$ terms in the continuity equation, else the harmonic charges cannot decay.
The isotropic contribution gives rise to the following terms:
\begin{equation}
    x^{i_1} x^{i_2} \cdots x^{i_n} \partial_a \partial^a J = 
    \partial_a (x^{i_1} x^{i_2} \cdots x^{i_n} \partial^a J -  n \delta^{a(i_1} x^{i_2} \cdots x^{i_n)} J) + 
    n(n-1) \delta^{(i_1 i_2} x^{i_3} \cdots x^{i_n)} J
    \, .
    \label{eqn:n-pole-continuity-scalar}
\end{equation}
It is simple to show from the expression~\eqref{eqn:n-pole-continuity-scalar} that the ``bulk'' contribution $n(n-1) \delta^{(i_1 i_2} x^{i_3} \cdots x^{i_n)} J $ vanishes for the harmonic polynomials in $d=2$.
Specifically, as before, we write the harmonic polynomials of degree $n$ as the real and imaginary parts of $(x + iy)^n = \sum_{k=0}^n \binom{n}{k} i^k y^k x^{n-k}$.
The source term in~\eqref{eqn:n-pole-continuity-scalar} then evaluates to
\begin{equation}
    J\sum_{k=0}^{n} \binom{n}{k} i^k x^{n-k} y^k \left[ (n-k)(n-k-1)x^{-2} + k(k-1) y^{-2}  \right]
    \, .
    \label{eqn:isotropic-bulk}
\end{equation}
By relabelling $k\to k+2$ in the second term in the square brackets, we observe that the two terms exactly cancel one another.
Meanwhile, for the other (quadrupole conserving) term in~\eqref{eqn:generic-n-pole-continuity} that couples to three spatial derivatives, we find that
\begin{multline}
    x^{i_1} x^{i_2} \cdots x^{i_n} \partial_a\partial_b \partial_c J^{abc} =\\ 
    \partial_a \left[x^{i_1}  \cdots x^{i_n} \partial_b \partial_c J^{abc} -  n x^{(i_1} \cdots x^{i_{n-1}} \partial_c J^{i_n)ac} + 
    n(n-1) x^{(i_1} \cdots x^{i_{n-2}} J^{i_{n-1} i_n ) a} \right] - n(n-1)(n-2) x^{(i_1} \cdots J^{i_{n-2} i_{n-1} i_n)}
    .
    \label{eqn:n-pole-continuity-tensor}
\end{multline}
For $n\geq 3$, there exists a source term.
It useful to write down the special cases for $n=0$ and $n=1$, corresponding to charge and dipole moment, respectively
\begin{gather}
    \partial_t \rho + \partial_a \left( \partial^a J + \partial_b \partial_c J^{abc} \right) = 0 \\
    \partial_t (x^\ell \rho) + \partial_a \left[  x^\ell \left(  \partial^a J + \partial_b \partial_c J^{abc} \right) -\partial_c J^{\ell a c} - \delta^{\ell a} J \right] = 0
    \, .
\end{gather}
These expressions allow us to identify the charge and dipole currents:
$J^a_\text{charge} = \partial^a J + \partial_b \partial_c J^{abc}$ and $J_\text{dipole}^{\ell a} = x^\ell J_\text{charge}^a - \partial_b J^{\ell a b} - \delta^{\ell a} J $, i.e., the dipole moment can change either due to the flow of charge across the boundary or due to the flow of dipoles.
In an analogous manner, the quadrupole current takes the form $J_\text{quad}^{\ell m a} = -x^\ell x^m J_\text{charge}^i + x^m J_\text{dipole}^{\ell a} + x^\ell J_\text{dipole}^{ma} + 2J^{\ell m a}$.
The boundary conditions relevant to the automaton circuits are zero flux of charge, dipole moment, and the $xy$ and $x^2 - y^2$ components of the quadrupole moment, which imply $n_a J_\text{charge}^a = 0$, $n_a(\partial_c J^{\ell a c} + \delta^{\ell a} J) = 0$, $n_a \sigma^x_{ij}J^{ij a} = 0$, and, finally, $n_a \sigma^z_{ij}J^{ij a} = 0$, where $\sigma_{ij}^{x/z}$ are the standard Pauli matrices.
Combining the isotropic term~\eqref{eqn:n-pole-continuity-scalar} and the anisotropic Eq.~\eqref{eqn:n-pole-continuity-tensor}, we arrive at the full continuity equation for the $n$-pole density
\begin{multline}
    \partial_t(x^{i_1} x^{i_2} \cdots x^{i_n} \rho) + 
    \partial_a (x^{i_1} x^{i_2} \cdots x^{i_n} [\overbrace{\partial^a J + \partial_b \partial_c J^{abc}}^{J_\text{charge}^a}] )
    - n\partial_a (   x^{(i_1} \cdots x^{i_{n-1}} [\overbrace{\delta^{i_n) a} J + \partial_c J^{i_n)ac}}^{x^{i_n)}J_\text{charge}^a- J_\text{dipole}^{i_n) a}} ] ) + \\
    \partial_a (n(n-1) x^{(i_1} \cdots x^{i_{n-2}} J^{i_{n-1} i_n ) a} ) +
    n(n-1) \delta^{(i_1 i_2} x^{i_3} \cdots x^{i_n)} J  -
    n(n-1)(n-2) x^{(i_1} \cdots J^{i_{n-2} i_{n-1} i_n)} = 0
    \, .
    \label{eqn:n-pole-continuity}
\end{multline}
The two terms with the overbraces are proportional to (linear combinations of) the charge and dipole current, implying that for zero flux boundary conditions they will not contribute.
In principle, the first term on the second line of~\eqref{eqn:n-pole-continuity} could give rise to a contribution proportional to $n_a \sigma^y_{ij} J^{ija}$ or $n_a \sigma^0_{ij} J^{ija}$, but we will show that such terms do not arise for the harmonic charges.
Expanding the complex harmonic charge, we find that
\begin{multline}
    n(n-1)\sum_{\{i_n\}} h(\{i_n\}) x^{(i_1} \cdots x^{i_{n-2}} J^{i_{n-1} i_n ) a} = \\
    \sum_{k=0}^n \binom{n}{k} i^k x^{n-k} y^k \left[
     k(k-1) y^{-2} J^{yya}
     +(n-k)(n-k-1) x^{-2} (x_2)^{k} J^{xxa}
     +2k(n-k) x^{-1} y^{-1} J^{xya}
    \right]
    \, ,
\end{multline}
where the coefficients $h(\{i_n\})$ correspond to the expansion of the $(x+iy)^n$.
As in Eq.~\eqref{eqn:isotropic-bulk}, we can relabel the dummy index $k$
and resum the expression, leading to
\begin{equation}
    n(n-1)\sum_{\{i_n\}} h(\{i_n\}) x^{(i_1} \cdots x^{i_{n-2}} J^{i_{n-1} i_n ) a} = n(n-1) (x+iy)^{n-2} \left[ (J^{xxa}-J^{yya})
     +2i
     J^{xya} \right]
     \, .
     \label{eqn:surface-term-n-pole}
\end{equation}
For boundary conditions that preserve $x^2-y^2$ and $xy$, i.e., $n_a J^{xya}=0$ and $n_a(J^{xxa} - J^{yya}) = 0$, the surface term corresponding to~\eqref{eqn:surface-term-n-pole} vanishes.
If follows that the only term that contributes when considering the time dependence of the harmonic charges is the last term on the bottom line of~\eqref{eqn:n-pole-continuity}, i.e., the anisotropic source term
\begin{equation}
    \frac{\mathrm{d}}{\mathrm{d}t} \mathcal{Q}_n
    =
    n(n-1)(n-2) \int \mathrm{d}^2 \mathbf{r} \; \left[ J^{xxx} + 3i J^{xxy} -3 J^{xyy} -iJ^{yyy} \right] (x + iy)^{n-3}
    \, .
\end{equation}
We may now immediately observe that $\mathcal{Q}_n$ is exactly conserved, even in the presence of boundaries, for $n \leq 2$ (by construction). Harmonic polynomials of degree $n >2$, on the other hand, are relaxed by the anisotropic quadrupole current tensor $J^{abc}$; the isotropic contribution to the equation does not enter the equation of motion for $\mathcal{Q}_n$.

\end{widetext}


\section{Solution of time-dependent biharmonic equation}
\label{sec:linear-exact-solution}

The Green's function for the isotropic theory can be expressed explicitly in terms of the hypergeometric function ${}_0F_2(a, b; x)$~\cite{abramovitz1964handbook} as 
\begin{equation}
G(\vec{r}; t) = \frac{1}{\sqrt{D_0t}} \left[
 \tfrac{1}{8\sqrt{\pi}}  \, _0F_2\left(\tfrac{1}{2},1; \tfrac{u^4}{4^4} \right)-\tfrac{u^2}{2\pi} \, _0F_2\left(\tfrac{3}{2},\tfrac{3}{2}; \tfrac{u^4}{4^4} \right) \right]
 \, ,
 \label{eqn:2d-greens-fn}
\end{equation}
where $u \equiv r(D_0t)^{-1/4}$, and $r \equiv |\vec{r}|$.
The Green's function of the one-dimensional problem, $\partial_t \rho + D_0 \partial_x^4 \rho = 0$, can be recovered by marginalising over one of the coordinate directions, $x$ or $y$, e.g., $G(x; t) = \int \mathrm{d}y \; G(\vec{r}; t)$. However, in contrast to ordinary diffusion, the two-dimensional Green's function in~\eqref{eqn:2d-greens-fn} is not simply the product of two one-dimensional solutions.

This exact solution can be used to evaluate the perturbative equation of motion for the charge $\mathcal{Q}_{xy}$ appearing in Eq.~\eqref{eq:isotropiccontinuumrelaxation}. Up to boundary terms,
\begin{subequations}
\begin{align}
    \frac{\mathrm{d}}{\mathrm{d}t} \mathcal{Q}_{xy} &= -2D_2 Q^3 \int
    \mathrm{d}^2 \vec{r} \; (\partial_x\partial_y G)^2 \partial_x^2 \partial_y^2 G \label{eqn:Qxy-eom-greens} \\
    &= \frac{1}{(D_0t)^3} I D_2 Q^3 
    \, ,
\end{align}
\end{subequations}
where $I > 0$ is the dimensionless integral that one obtains by introducing dimensionless integration variables $u_i = x_i (D_0t)^{-1/4}$ in~\eqref{eqn:Qxy-eom-greens}. Numerically, we find that the integral evaluates to $I = 6.34918\ldots \times 10^{-7}$. This means that a positive (negative) $D_2$ will, asymptotically, give rise to algebraic-in-time growth (decay) of the harmonic charge $\mathcal{Q}_{xy}$.


\section{Hamiltonians that conserve traceless components of quadrupole}
\label{sec:local-hamiltonians}

In the outlook, we introduced two local spin Hamiltonians that were constructed from local ``hopping'' terms that conserve $xy$ and $x^2 - y^2$, in addition to charge, $\sum_i \hat{S}_i^z$, and dipole moment, $\sum_i x_i^\ell \hat{S}_i^z$ for $\ell = x, y$.
It was asserted that these Hamiltonians \emph{also} conserve lattice versions of the harmonic charges. We show here how these additional conservation laws emerge.
For concreteness, we repeat the definitions (where we have set the overall energy scale $J=1$)
\begin{subequations}
\begin{align}
    \hat{H}^{(3)} &= \sum_\vec{r} \left( \prod_{\boldsymbol{\delta}\in \{\pm {\vec{e}}_x, \pm \vec{e}_y\}} \hat{S}_{\vec{r}}^+ \hat{S}_{\vec{r}+\boldsymbol{\delta}}^- + \text{H.c.} \right)
    \, , \\
    \hat{H}^{(5)} &= \sum_\vec{r} \left( \prod_{\boldsymbol{\delta}\in \{\pm \vec{e}_x, \pm \vec{e}_y\}} \hat{S}_{\vec{r}+\boldsymbol{\delta}}^+ \hat{S}_{\vec{r}+2\boldsymbol{\delta}}^- + \text{H.c.} \right)
    \, .
\end{align}
\end{subequations}
Consider the Heisenberg equation of motion for the operator $\hat{\mathcal{Q}}_f = \sum_i f(\vec{r}_i) \hat{S}_i^z$.
Making use of the commutator
$[\hat{S}_i^z, \hat{S}_j^\pm] =\pm \delta_{ij} \hat{S}_i^\pm$, which implies that $[\hat{S}_i^z, (\hat{S}_j^\pm)^n] = \pm n (\hat{S}_i^\pm)^n$, the commutator $[\hat{\mathcal{Q}}_f, \hat{H}]$ that determines the time evolution of $\hat{\mathcal{Q}}_f$ can be evaluated.
For $\hat{H}^{(3)}$, one finds that
\begin{equation}
    [\hat{\mathcal{Q}}_f, \hat{H}^{(3)}] = 
    \sum_\vec{r} \left( 4f_{\vec{r}} - \sum_{\boldsymbol{\delta}} f_{\vec{r}+\boldsymbol{\delta}} \right) (\prod_\vec{\boldsymbol{\delta}} \hat{S}_{\vec{r}}^+ \hat{S}_{\vec{r}+\boldsymbol{\delta}}^- - \text{H.c.})
    \, ,
\end{equation}
while for the Hamiltonian $\hat{H}^{(5)}$, constructed from a larger stencil,
\begin{equation}
    [\hat{\mathcal{Q}}_f, \hat{H}^{(5)}] =
    \sum_{\mathbf{r}, \boldsymbol{\delta}}
    \left( f_{\vec{r}+\boldsymbol{\delta}} -  f_{\vec{r}+2\boldsymbol{\delta}} \right) (\prod_\vec{\boldsymbol{\delta}} \hat{S}_{\mathbf{r}+\boldsymbol{\delta}}^+ \hat{S}_{\mathbf{r}+2\boldsymbol{\delta}}^- - \text{H.c.})
    \, .
\end{equation}
Any moment defined by a discrete function $f(\vec{r}) \equiv f_\vec{r}$ that satisfies
\begin{equation}
   4f_{\vec{r}} - \sum_{\boldsymbol{\delta}} f_{\vec{r}+\boldsymbol{\delta}} = 0
    \quad\text{or}\quad
    \sum_{\boldsymbol{\delta}}
    \left( f_{\vec{r}+\boldsymbol{\delta}} -  f_{\vec{r}+2\boldsymbol{\delta}} \right) = 0
    \, ,
    \label{eqn:discrete-laplace}
\end{equation}
for $\hat{H}^{(3)}$ or $\hat{H}^{(5)}$, respectively, for all $\vec{r}$ is exactly conserved, up to potential boundary terms in the Hamiltonian.
Both of these equations are discretised versions of the two-dimensional Laplace equation, i.e., each of the equations may be written as $\sum_{\vec{r}'} \Delta_{\vec{r}\vec{r}'} f_{\vec{r}'} = 0$, where $\Delta_{\vec{r}\vec{r}'}$ corresponds to a finite difference approximation to the two-dimensional Laplacian $\laplacian$. Both $\hat{H}^{(3)}$ and $\hat{H}^{(5)}$ lead to second-order central finite difference approximations to $\laplacian$; in principle, Hamiltonians involving additional ``hopping'' terms may be constructed that act as higher-order central finite difference operators.

Is is easy to verify that the functions $f = \{1, x, y, xy, x^2 - y^2\}$ satisfy exactly both equations in~\eqref{eqn:discrete-laplace}, as required. Some of the lowest-degree harmonic \emph{polynomials} are also exactly conserved, e.g., $f=\{x^3-3xy^2, 3x^2y-y^3,x^3y-xy^3\}$. More generally, the discrete Laplace equation can be inverted for a given $\{ f_\vec{r} \, | \, \vec{r} \in \partial \}$ (i.e., $f_\vec{r}$ on the boundary). The number of linearly independent solutions therefore scales with the number of boundary sites.


\bibliography{library}

\begin{thebibliography}{42}%
\makeatletter
\providecommand \@ifxundefined [1]{%
 \@ifx{#1\undefined}
}%
\providecommand \@ifnum [1]{%
 \ifnum #1\expandafter \@firstoftwo
 \else \expandafter \@secondoftwo
 \fi
}%
\providecommand \@ifx [1]{%
 \ifx #1\expandafter \@firstoftwo
 \else \expandafter \@secondoftwo
 \fi
}%
\providecommand \natexlab [1]{#1}%
\providecommand \enquote  [1]{``#1''}%
\providecommand \bibnamefont  [1]{#1}%
\providecommand \bibfnamefont [1]{#1}%
\providecommand \citenamefont [1]{#1}%
\providecommand \href@noop [0]{\@secondoftwo}%
\providecommand \href [0]{\begingroup \@sanitize@url \@href}%
\providecommand \@href[1]{\@@startlink{#1}\@@href}%
\providecommand \@@href[1]{\endgroup#1\@@endlink}%
\providecommand \@sanitize@url [0]{\catcode `\\12\catcode `\$12\catcode
  `\&12\catcode `\#12\catcode `\^12\catcode `\_12\catcode `\%12\relax}%
\providecommand \@@startlink[1]{}%
\providecommand \@@endlink[0]{}%
\providecommand \url  [0]{\begingroup\@sanitize@url \@url }%
\providecommand \@url [1]{\endgroup\@href {#1}{\urlprefix }}%
\providecommand \urlprefix  [0]{URL }%
\providecommand \Eprint [0]{\href }%
\providecommand \doibase [0]{https://doi.org/}%
\providecommand \selectlanguage [0]{\@gobble}%
\providecommand \bibinfo  [0]{\@secondoftwo}%
\providecommand \bibfield  [0]{\@secondoftwo}%
\providecommand \translation [1]{[#1]}%
\providecommand \BibitemOpen [0]{}%
\providecommand \bibitemStop [0]{}%
\providecommand \bibitemNoStop [0]{.\EOS\space}%
\providecommand \EOS [0]{\spacefactor3000\relax}%
\providecommand \BibitemShut  [1]{\csname bibitem#1\endcsname}%
\let\auto@bib@innerbib\@empty
\bibitem [{\citenamefont {Pai}\ \emph {et~al.}(2019)\citenamefont {Pai},
  \citenamefont {Pretko},\ and\ \citenamefont {Nandkishore}}]{PPN}%
  \BibitemOpen
  \bibfield  {author} {\bibinfo {author} {\bibfnamefont {S.}~\bibnamefont
  {Pai}}, \bibinfo {author} {\bibfnamefont {M.}~\bibnamefont {Pretko}},\ and\
  \bibinfo {author} {\bibfnamefont {R.~M.}\ \bibnamefont {Nandkishore}},\
  }\bibfield  {title} {\bibinfo {title} {Localization in fractonic random
  circuits},\ }\href {https://doi.org/10.1103/PhysRevX.9.021003} {\bibfield
  {journal} {\bibinfo  {journal} {Phys. Rev. X}\ }\textbf {\bibinfo {volume}
  {9}},\ \bibinfo {pages} {021003} (\bibinfo {year} {2019})}\BibitemShut
  {NoStop}%
\bibitem [{\citenamefont {Khemani}\ \emph {et~al.}(2020)\citenamefont
  {Khemani}, \citenamefont {Hermele},\ and\ \citenamefont
  {Nandkishore}}]{KhemaniShattering}%
  \BibitemOpen
  \bibfield  {author} {\bibinfo {author} {\bibfnamefont {V.}~\bibnamefont
  {Khemani}}, \bibinfo {author} {\bibfnamefont {M.}~\bibnamefont {Hermele}},\
  and\ \bibinfo {author} {\bibfnamefont {R.}~\bibnamefont {Nandkishore}},\
  }\bibfield  {title} {\bibinfo {title} {Localization from hilbert space
  shattering: From theory to physical realizations},\ }\href
  {https://doi.org/10.1103/PhysRevB.101.174204} {\bibfield  {journal} {\bibinfo
   {journal} {Phys. Rev. B}\ }\textbf {\bibinfo {volume} {101}},\ \bibinfo
  {pages} {174204} (\bibinfo {year} {2020})}\BibitemShut {NoStop}%
\bibitem [{\citenamefont {Sala}\ \emph {et~al.}(2020)\citenamefont {Sala},
  \citenamefont {Rakovszky}, \citenamefont {Verresen}, \citenamefont {Knap},\
  and\ \citenamefont {Pollmann}}]{SalaFragmentation}%
  \BibitemOpen
  \bibfield  {author} {\bibinfo {author} {\bibfnamefont {P.}~\bibnamefont
  {Sala}}, \bibinfo {author} {\bibfnamefont {T.}~\bibnamefont {Rakovszky}},
  \bibinfo {author} {\bibfnamefont {R.}~\bibnamefont {Verresen}}, \bibinfo
  {author} {\bibfnamefont {M.}~\bibnamefont {Knap}},\ and\ \bibinfo {author}
  {\bibfnamefont {F.}~\bibnamefont {Pollmann}},\ }\bibfield  {title} {\bibinfo
  {title} {Ergodicity breaking arising from hilbert space fragmentation in
  dipole-conserving hamiltonians},\ }\href
  {https://doi.org/10.1103/PhysRevX.10.011047} {\bibfield  {journal} {\bibinfo
  {journal} {Phys. Rev. X}\ }\textbf {\bibinfo {volume} {10}},\ \bibinfo
  {pages} {011047} (\bibinfo {year} {2020})}\BibitemShut {NoStop}%
\bibitem [{\citenamefont {Moudgalya}\ \emph {et~al.}(2021)\citenamefont
  {Moudgalya}, \citenamefont {Prem}, \citenamefont {Nandkishore}, \citenamefont
  {Regnault},\ and\ \citenamefont {Bernevig}}]{moudgalyaprem}%
  \BibitemOpen
  \bibfield  {author} {\bibinfo {author} {\bibfnamefont {S.}~\bibnamefont
  {Moudgalya}}, \bibinfo {author} {\bibfnamefont {A.}~\bibnamefont {Prem}},
  \bibinfo {author} {\bibfnamefont {R.}~\bibnamefont {Nandkishore}}, \bibinfo
  {author} {\bibfnamefont {N.}~\bibnamefont {Regnault}},\ and\ \bibinfo
  {author} {\bibfnamefont {B.~A.}\ \bibnamefont {Bernevig}},\ }\bibfield
  {title} {\bibinfo {title} {Thermalization and its absence within krylov
  subspaces of a constrained hamiltonian},\ }in\ \href
  {https://doi.org/10.1142/9789811231711_0009} {\emph {\bibinfo {booktitle}
  {Memorial Volume for Shoucheng Zhang}}}\ (\bibinfo  {publisher} {WORLD
  SCIENTIFIC},\ \bibinfo {year} {2021})\ Chap.\ \bibinfo {chapter} {Chapter 7},
  pp.\ \bibinfo {pages} {147--209}\BibitemShut {NoStop}%
\bibitem [{\citenamefont {Rakovszky}\ \emph {et~al.}(2020)\citenamefont
  {Rakovszky}, \citenamefont {Sala}, \citenamefont {Verresen}, \citenamefont
  {Knap},\ and\ \citenamefont {Pollmann}}]{SLIOMs}%
  \BibitemOpen
  \bibfield  {author} {\bibinfo {author} {\bibfnamefont {T.}~\bibnamefont
  {Rakovszky}}, \bibinfo {author} {\bibfnamefont {P.}~\bibnamefont {Sala}},
  \bibinfo {author} {\bibfnamefont {R.}~\bibnamefont {Verresen}}, \bibinfo
  {author} {\bibfnamefont {M.}~\bibnamefont {Knap}},\ and\ \bibinfo {author}
  {\bibfnamefont {F.}~\bibnamefont {Pollmann}},\ }\bibfield  {title} {\bibinfo
  {title} {Statistical localization: From strong fragmentation to strong edge
  modes},\ }\href {https://doi.org/10.1103/PhysRevB.101.125126} {\bibfield
  {journal} {\bibinfo  {journal} {Phys. Rev. B}\ }\textbf {\bibinfo {volume}
  {101}},\ \bibinfo {pages} {125126} (\bibinfo {year} {2020})}\BibitemShut
  {NoStop}%
\bibitem [{\citenamefont {Moudgalya}\ and\ \citenamefont
  {Motrunich}(2022)}]{moudgalyamotrunich}%
  \BibitemOpen
  \bibfield  {author} {\bibinfo {author} {\bibfnamefont {S.}~\bibnamefont
  {Moudgalya}}\ and\ \bibinfo {author} {\bibfnamefont {O.~I.}\ \bibnamefont
  {Motrunich}},\ }\bibfield  {title} {\bibinfo {title} {Hilbert space
  fragmentation and commutant algebras},\ }\href
  {https://doi.org/10.1103/PhysRevX.12.011050} {\bibfield  {journal} {\bibinfo
  {journal} {Phys. Rev. X}\ }\textbf {\bibinfo {volume} {12}},\ \bibinfo
  {pages} {011050} (\bibinfo {year} {2022})}\BibitemShut {NoStop}%
\bibitem [{\citenamefont {Khudorozhkov}\ \emph {et~al.}(2021)\citenamefont
  {Khudorozhkov}, \citenamefont {Tiwari}, \citenamefont {Chamon},\ and\
  \citenamefont {Neupert}}]{khudorozhkov2021hilbert}%
  \BibitemOpen
  \bibfield  {author} {\bibinfo {author} {\bibfnamefont {A.}~\bibnamefont
  {Khudorozhkov}}, \bibinfo {author} {\bibfnamefont {A.}~\bibnamefont
  {Tiwari}}, \bibinfo {author} {\bibfnamefont {C.}~\bibnamefont {Chamon}},\
  and\ \bibinfo {author} {\bibfnamefont {T.}~\bibnamefont {Neupert}},\
  }\href@noop {} {\bibinfo {title} {Hilbert space fragmentation in a 2d quantum
  spin system with subsystem symmetries}} (\bibinfo {year} {2021}),\ \Eprint
  {https://arxiv.org/abs/2107.09690} {arXiv:2107.09690 [cond-mat.str-el]}
  \BibitemShut {NoStop}%
\bibitem [{\citenamefont {Iaconis}\ \emph {et~al.}(2019)\citenamefont
  {Iaconis}, \citenamefont {Vijay},\ and\ \citenamefont
  {Nandkishore}}]{IaconisSubsystem2019}%
  \BibitemOpen
  \bibfield  {author} {\bibinfo {author} {\bibfnamefont {J.}~\bibnamefont
  {Iaconis}}, \bibinfo {author} {\bibfnamefont {S.}~\bibnamefont {Vijay}},\
  and\ \bibinfo {author} {\bibfnamefont {R.}~\bibnamefont {Nandkishore}},\
  }\bibfield  {title} {\bibinfo {title} {Anomalous subdiffusion from subsystem
  symmetries},\ }\href {https://doi.org/10.1103/PhysRevB.100.214301} {\bibfield
   {journal} {\bibinfo  {journal} {Phys. Rev. B}\ }\textbf {\bibinfo {volume}
  {100}},\ \bibinfo {pages} {214301} (\bibinfo {year} {2019})}\BibitemShut
  {NoStop}%
\bibitem [{\citenamefont {Gromov}\ \emph {et~al.}(2020)\citenamefont {Gromov},
  \citenamefont {Lucas},\ and\ \citenamefont {Nandkishore}}]{fractonhydro}%
  \BibitemOpen
  \bibfield  {author} {\bibinfo {author} {\bibfnamefont {A.}~\bibnamefont
  {Gromov}}, \bibinfo {author} {\bibfnamefont {A.}~\bibnamefont {Lucas}},\ and\
  \bibinfo {author} {\bibfnamefont {R.~M.}\ \bibnamefont {Nandkishore}},\
  }\bibfield  {title} {\bibinfo {title} {Fracton hydrodynamics},\ }\href
  {https://doi.org/10.1103/PhysRevResearch.2.033124} {\bibfield  {journal}
  {\bibinfo  {journal} {Phys. Rev. Research}\ }\textbf {\bibinfo {volume}
  {2}},\ \bibinfo {pages} {033124} (\bibinfo {year} {2020})}\BibitemShut
  {NoStop}%
\bibitem [{\citenamefont {Feldmeier}\ \emph {et~al.}(2020)\citenamefont
  {Feldmeier}, \citenamefont {Sala}, \citenamefont {De~Tomasi}, \citenamefont
  {Pollmann},\ and\ \citenamefont {Knap}}]{FeldmeierAnomalous2020}%
  \BibitemOpen
  \bibfield  {author} {\bibinfo {author} {\bibfnamefont {J.}~\bibnamefont
  {Feldmeier}}, \bibinfo {author} {\bibfnamefont {P.}~\bibnamefont {Sala}},
  \bibinfo {author} {\bibfnamefont {G.}~\bibnamefont {De~Tomasi}}, \bibinfo
  {author} {\bibfnamefont {F.}~\bibnamefont {Pollmann}},\ and\ \bibinfo
  {author} {\bibfnamefont {M.}~\bibnamefont {Knap}},\ }\bibfield  {title}
  {\bibinfo {title} {Anomalous diffusion in dipole- and
  higher-moment-conserving systems},\ }\href
  {https://doi.org/10.1103/PhysRevLett.125.245303} {\bibfield  {journal}
  {\bibinfo  {journal} {Phys. Rev. Lett.}\ }\textbf {\bibinfo {volume} {125}},\
  \bibinfo {pages} {245303} (\bibinfo {year} {2020})}\BibitemShut {NoStop}%
\bibitem [{\citenamefont {Morningstar}\ \emph {et~al.}(2020)\citenamefont
  {Morningstar}, \citenamefont {Khemani},\ and\ \citenamefont
  {Huse}}]{Morningstar}%
  \BibitemOpen
  \bibfield  {author} {\bibinfo {author} {\bibfnamefont {A.}~\bibnamefont
  {Morningstar}}, \bibinfo {author} {\bibfnamefont {V.}~\bibnamefont
  {Khemani}},\ and\ \bibinfo {author} {\bibfnamefont {D.~A.}\ \bibnamefont
  {Huse}},\ }\bibfield  {title} {\bibinfo {title} {Kinetically constrained
  freezing transition in a dipole-conserving system},\ }\href
  {https://doi.org/10.1103/PhysRevB.101.214205} {\bibfield  {journal} {\bibinfo
   {journal} {Phys. Rev. B}\ }\textbf {\bibinfo {volume} {101}},\ \bibinfo
  {pages} {214205} (\bibinfo {year} {2020})}\BibitemShut {NoStop}%
\bibitem [{\citenamefont {Iaconis}\ \emph {et~al.}(2021)\citenamefont
  {Iaconis}, \citenamefont {Lucas},\ and\ \citenamefont
  {Nandkishore}}]{IaconisAnyDimension2021}%
  \BibitemOpen
  \bibfield  {author} {\bibinfo {author} {\bibfnamefont {J.}~\bibnamefont
  {Iaconis}}, \bibinfo {author} {\bibfnamefont {A.}~\bibnamefont {Lucas}},\
  and\ \bibinfo {author} {\bibfnamefont {R.}~\bibnamefont {Nandkishore}},\
  }\bibfield  {title} {\bibinfo {title} {Multipole conservation laws and
  subdiffusion in any dimension},\ }\href
  {https://doi.org/10.1103/PhysRevE.103.022142} {\bibfield  {journal} {\bibinfo
   {journal} {Phys. Rev. E}\ }\textbf {\bibinfo {volume} {103}},\ \bibinfo
  {pages} {022142} (\bibinfo {year} {2021})}\BibitemShut {NoStop}%
\bibitem [{\citenamefont {Glorioso}\ \emph {et~al.}(2021)\citenamefont
  {Glorioso}, \citenamefont {Guo}, \citenamefont {Rodriguez-Nieva},\ and\
  \citenamefont {Lucas}}]{Glorioso}%
  \BibitemOpen
  \bibfield  {author} {\bibinfo {author} {\bibfnamefont {P.}~\bibnamefont
  {Glorioso}}, \bibinfo {author} {\bibfnamefont {J.}~\bibnamefont {Guo}},
  \bibinfo {author} {\bibfnamefont {J.~F.}\ \bibnamefont {Rodriguez-Nieva}},\
  and\ \bibinfo {author} {\bibfnamefont {A.}~\bibnamefont {Lucas}},\
  }\href@noop {} {\bibinfo {title} {Breakdown of hydrodynamics below four
  dimensions in a fracton fluid}} (\bibinfo {year} {2021}),\ \Eprint
  {https://arxiv.org/abs/2105.13365} {arXiv:2105.13365 [cond-mat.str-el]}
  \BibitemShut {NoStop}%
\bibitem [{\citenamefont {Doshi}\ and\ \citenamefont
  {Gromov}(2021)}]{Doshi:2020jso}%
  \BibitemOpen
  \bibfield  {author} {\bibinfo {author} {\bibfnamefont {D.}~\bibnamefont
  {Doshi}}\ and\ \bibinfo {author} {\bibfnamefont {A.}~\bibnamefont {Gromov}},\
  }\bibfield  {title} {\bibinfo {title} {Vortices as fractons},\ }\href
  {https://doi.org/10.1038/s42005-021-00540-4} {\bibfield  {journal} {\bibinfo
  {journal} {Communications Physics}\ }\textbf {\bibinfo {volume} {4}},\
  \bibinfo {pages} {1} (\bibinfo {year} {2021})}\BibitemShut {NoStop}%
\bibitem [{\citenamefont {Grosvenor}\ \emph {et~al.}(2021)\citenamefont
  {Grosvenor}, \citenamefont {Hoyos}, \citenamefont {Pe\~na Benitez},\ and\
  \citenamefont {Sur\'owka}}]{Grosvenor:2021rrt}%
  \BibitemOpen
  \bibfield  {author} {\bibinfo {author} {\bibfnamefont {K.~T.}\ \bibnamefont
  {Grosvenor}}, \bibinfo {author} {\bibfnamefont {C.}~\bibnamefont {Hoyos}},
  \bibinfo {author} {\bibfnamefont {F.}~\bibnamefont {Pe\~na Benitez}},\ and\
  \bibinfo {author} {\bibfnamefont {P.}~\bibnamefont {Sur\'owka}},\ }\bibfield
  {title} {\bibinfo {title} {Hydrodynamics of ideal fracton fluids},\ }\href
  {https://doi.org/10.1103/PhysRevResearch.3.043186} {\bibfield  {journal}
  {\bibinfo  {journal} {Phys. Rev. Research}\ }\textbf {\bibinfo {volume}
  {3}},\ \bibinfo {pages} {043186} (\bibinfo {year} {2021})}\BibitemShut
  {NoStop}%
\bibitem [{\citenamefont {Guardado-Sanchez}\ \emph {et~al.}(2020)\citenamefont
  {Guardado-Sanchez}, \citenamefont {Morningstar}, \citenamefont {Spar},
  \citenamefont {Brown}, \citenamefont {Huse},\ and\ \citenamefont
  {Bakr}}]{Bakr}%
  \BibitemOpen
  \bibfield  {author} {\bibinfo {author} {\bibfnamefont {E.}~\bibnamefont
  {Guardado-Sanchez}}, \bibinfo {author} {\bibfnamefont {A.}~\bibnamefont
  {Morningstar}}, \bibinfo {author} {\bibfnamefont {B.~M.}\ \bibnamefont
  {Spar}}, \bibinfo {author} {\bibfnamefont {P.~T.}\ \bibnamefont {Brown}},
  \bibinfo {author} {\bibfnamefont {D.~A.}\ \bibnamefont {Huse}},\ and\
  \bibinfo {author} {\bibfnamefont {W.~S.}\ \bibnamefont {Bakr}},\ }\bibfield
  {title} {\bibinfo {title} {Subdiffusion and heat transport in a tilted
  two-dimensional fermi-hubbard system},\ }\href
  {https://doi.org/10.1103/PhysRevX.10.011042} {\bibfield  {journal} {\bibinfo
  {journal} {Phys. Rev. X}\ }\textbf {\bibinfo {volume} {10}},\ \bibinfo
  {pages} {011042} (\bibinfo {year} {2020})}\BibitemShut {NoStop}%
\bibitem [{\citenamefont {Scherg}\ \emph {et~al.}(2021)\citenamefont {Scherg},
  \citenamefont {Kohlert}, \citenamefont {Sala}, \citenamefont {Pollmann},
  \citenamefont {Madhusudhana}, \citenamefont {Bloch},\ and\ \citenamefont
  {Aidelsburger}}]{Scherg2021observing}%
  \BibitemOpen
  \bibfield  {author} {\bibinfo {author} {\bibfnamefont {S.}~\bibnamefont
  {Scherg}}, \bibinfo {author} {\bibfnamefont {T.}~\bibnamefont {Kohlert}},
  \bibinfo {author} {\bibfnamefont {P.}~\bibnamefont {Sala}}, \bibinfo {author}
  {\bibfnamefont {F.}~\bibnamefont {Pollmann}}, \bibinfo {author}
  {\bibfnamefont {B.~H.}\ \bibnamefont {Madhusudhana}}, \bibinfo {author}
  {\bibfnamefont {I.}~\bibnamefont {Bloch}},\ and\ \bibinfo {author}
  {\bibfnamefont {M.}~\bibnamefont {Aidelsburger}},\ }\bibfield  {title}
  {\bibinfo {title} {Observing non-ergodicity due to kinetic constraints in
  tilted fermi-hubbard chains},\ }\href
  {https://doi.org/https://doi.org/10.1038/s41467-021-24726-0} {\bibfield
  {journal} {\bibinfo  {journal} {Nature Communications}\ }\textbf {\bibinfo
  {volume} {12}},\ \bibinfo {pages} {1} (\bibinfo {year} {2021})}\BibitemShut
  {NoStop}%
\bibitem [{\citenamefont {Kohlert}\ \emph {et~al.}(2021)\citenamefont
  {Kohlert}, \citenamefont {Scherg}, \citenamefont {Sala}, \citenamefont
  {Pollmann}, \citenamefont {Madhusudhana}, \citenamefont {Bloch},\ and\
  \citenamefont {Aidelsburger}}]{Kohlert2021experimental}%
  \BibitemOpen
  \bibfield  {author} {\bibinfo {author} {\bibfnamefont {T.}~\bibnamefont
  {Kohlert}}, \bibinfo {author} {\bibfnamefont {S.}~\bibnamefont {Scherg}},
  \bibinfo {author} {\bibfnamefont {P.}~\bibnamefont {Sala}}, \bibinfo {author}
  {\bibfnamefont {F.}~\bibnamefont {Pollmann}}, \bibinfo {author}
  {\bibfnamefont {B.~H.}\ \bibnamefont {Madhusudhana}}, \bibinfo {author}
  {\bibfnamefont {I.}~\bibnamefont {Bloch}},\ and\ \bibinfo {author}
  {\bibfnamefont {M.}~\bibnamefont {Aidelsburger}},\ }\href@noop {} {\bibinfo
  {title} {Experimental realization of fragmented models in tilted
  fermi-hubbard chains}} (\bibinfo {year} {2021}),\ \Eprint
  {https://arxiv.org/abs/2106.15586} {arXiv:2106.15586 [cond-mat.quant-gas]}
  \BibitemShut {NoStop}%
\bibitem [{\citenamefont {Guo}\ \emph {et~al.}(2021)\citenamefont {Guo},
  \citenamefont {Cheng}, \citenamefont {Li}, \citenamefont {Xu}, \citenamefont
  {Zhang}, \citenamefont {Wang}, \citenamefont {Song}, \citenamefont {Liu},
  \citenamefont {Ren}, \citenamefont {Dong}, \citenamefont {Mondaini},\ and\
  \citenamefont {Wang}}]{Guo}%
  \BibitemOpen
  \bibfield  {author} {\bibinfo {author} {\bibfnamefont {Q.}~\bibnamefont
  {Guo}}, \bibinfo {author} {\bibfnamefont {C.}~\bibnamefont {Cheng}}, \bibinfo
  {author} {\bibfnamefont {H.}~\bibnamefont {Li}}, \bibinfo {author}
  {\bibfnamefont {S.}~\bibnamefont {Xu}}, \bibinfo {author} {\bibfnamefont
  {P.}~\bibnamefont {Zhang}}, \bibinfo {author} {\bibfnamefont
  {Z.}~\bibnamefont {Wang}}, \bibinfo {author} {\bibfnamefont {C.}~\bibnamefont
  {Song}}, \bibinfo {author} {\bibfnamefont {W.}~\bibnamefont {Liu}}, \bibinfo
  {author} {\bibfnamefont {W.}~\bibnamefont {Ren}}, \bibinfo {author}
  {\bibfnamefont {H.}~\bibnamefont {Dong}}, \bibinfo {author} {\bibfnamefont
  {R.}~\bibnamefont {Mondaini}},\ and\ \bibinfo {author} {\bibfnamefont
  {H.}~\bibnamefont {Wang}},\ }\bibfield  {title} {\bibinfo {title} {Stark
  many-body localization on a superconducting quantum processor},\ }\href
  {https://doi.org/10.1103/PhysRevLett.127.240502} {\bibfield  {journal}
  {\bibinfo  {journal} {Phys. Rev. Lett.}\ }\textbf {\bibinfo {volume} {127}},\
  \bibinfo {pages} {240502} (\bibinfo {year} {2021})}\BibitemShut {NoStop}%
\bibitem [{\citenamefont
  {Pretko}(2017{\natexlab{a}})}]{PretkoSubdimensional2017}%
  \BibitemOpen
  \bibfield  {author} {\bibinfo {author} {\bibfnamefont {M.}~\bibnamefont
  {Pretko}},\ }\bibfield  {title} {\bibinfo {title} {{Subdimensional particle
  structure of higher rank $U(1)$ spin liquids}},\ }\href
  {https://doi.org/10.1103/PhysRevB.95.115139} {\bibfield  {journal} {\bibinfo
  {journal} {Phys. Rev. B}\ }\textbf {\bibinfo {volume} {95}},\ \bibinfo
  {pages} {115139} (\bibinfo {year} {2017}{\natexlab{a}})}\BibitemShut
  {NoStop}%
\bibitem [{\citenamefont {Chamon}(2005)}]{ChamonQuantumGlassiness2005}%
  \BibitemOpen
  \bibfield  {author} {\bibinfo {author} {\bibfnamefont {C.}~\bibnamefont
  {Chamon}},\ }\bibfield  {title} {\bibinfo {title} {Quantum glassiness in
  strongly correlated clean systems: An example of topological
  overprotection},\ }\href {https://doi.org/10.1103/PhysRevLett.94.040402}
  {\bibfield  {journal} {\bibinfo  {journal} {Phys. Rev. Lett.}\ }\textbf
  {\bibinfo {volume} {94}},\ \bibinfo {pages} {040402} (\bibinfo {year}
  {2005})}\BibitemShut {NoStop}%
\bibitem [{\citenamefont {Haah}(2011)}]{Haah2011}%
  \BibitemOpen
  \bibfield  {author} {\bibinfo {author} {\bibfnamefont {J.}~\bibnamefont
  {Haah}},\ }\bibfield  {title} {\bibinfo {title} {Local stabilizer codes in
  three dimensions without string logical operators},\ }\href
  {https://doi.org/10.1103/PhysRevA.83.042330} {\bibfield  {journal} {\bibinfo
  {journal} {Phys. Rev. A}\ }\textbf {\bibinfo {volume} {83}},\ \bibinfo
  {pages} {042330} (\bibinfo {year} {2011})}\BibitemShut {NoStop}%
\bibitem [{\citenamefont {Vijay}\ \emph {et~al.}(2015)\citenamefont {Vijay},
  \citenamefont {Haah},\ and\ \citenamefont {Fu}}]{Vijay2015}%
  \BibitemOpen
  \bibfield  {author} {\bibinfo {author} {\bibfnamefont {S.}~\bibnamefont
  {Vijay}}, \bibinfo {author} {\bibfnamefont {J.}~\bibnamefont {Haah}},\ and\
  \bibinfo {author} {\bibfnamefont {L.}~\bibnamefont {Fu}},\ }\bibfield
  {title} {\bibinfo {title} {A new kind of topological quantum order: A
  dimensional hierarchy of quasiparticles built from stationary excitations},\
  }\href {https://doi.org/10.1103/PhysRevB.92.235136} {\bibfield  {journal}
  {\bibinfo  {journal} {Phys. Rev. B}\ }\textbf {\bibinfo {volume} {92}},\
  \bibinfo {pages} {235136} (\bibinfo {year} {2015})}\BibitemShut {NoStop}%
\bibitem [{\citenamefont {Vijay}\ \emph {et~al.}(2016)\citenamefont {Vijay},
  \citenamefont {Haah},\ and\ \citenamefont {Fu}}]{VijayTopoOrder2016}%
  \BibitemOpen
  \bibfield  {author} {\bibinfo {author} {\bibfnamefont {S.}~\bibnamefont
  {Vijay}}, \bibinfo {author} {\bibfnamefont {J.}~\bibnamefont {Haah}},\ and\
  \bibinfo {author} {\bibfnamefont {L.}~\bibnamefont {Fu}},\ }\bibfield
  {title} {\bibinfo {title} {Fracton topological order, generalized lattice
  gauge theory, and duality},\ }\href
  {https://doi.org/10.1103/PhysRevB.94.235157} {\bibfield  {journal} {\bibinfo
  {journal} {Phys. Rev. B}\ }\textbf {\bibinfo {volume} {94}},\ \bibinfo
  {pages} {235157} (\bibinfo {year} {2016})}\BibitemShut {NoStop}%
\bibitem [{\citenamefont {Nandkishore}\ and\ \citenamefont
  {Hermele}(2019)}]{NandkishoreFractonsAnnurev}%
  \BibitemOpen
  \bibfield  {author} {\bibinfo {author} {\bibfnamefont {R.~M.}\ \bibnamefont
  {Nandkishore}}\ and\ \bibinfo {author} {\bibfnamefont {M.}~\bibnamefont
  {Hermele}},\ }\bibfield  {title} {\bibinfo {title} {Fractons},\ }\href
  {https://doi.org/10.1146/annurev-conmatphys-031218-013604} {\bibfield
  {journal} {\bibinfo  {journal} {Annual Review of Condensed Matter Physics}\
  }\textbf {\bibinfo {volume} {10}},\ \bibinfo {pages} {295} (\bibinfo {year}
  {2019})}\BibitemShut {NoStop}%
\bibitem [{\citenamefont {Pretko}(2017{\natexlab{b}})}]{PretkoGeneralizedEM}%
  \BibitemOpen
  \bibfield  {author} {\bibinfo {author} {\bibfnamefont {M.}~\bibnamefont
  {Pretko}},\ }\bibfield  {title} {\bibinfo {title} {Generalized
  electromagnetism of subdimensional particles: A spin liquid story},\ }\href
  {https://doi.org/10.1103/PhysRevB.96.035119} {\bibfield  {journal} {\bibinfo
  {journal} {Phys. Rev. B}\ }\textbf {\bibinfo {volume} {96}},\ \bibinfo
  {pages} {035119} (\bibinfo {year} {2017}{\natexlab{b}})}\BibitemShut
  {NoStop}%
\bibitem [{\citenamefont {Bulmash}\ and\ \citenamefont
  {Barkeshli}(2018)}]{BulmashBarkeshli2018}%
  \BibitemOpen
  \bibfield  {author} {\bibinfo {author} {\bibfnamefont {D.}~\bibnamefont
  {Bulmash}}\ and\ \bibinfo {author} {\bibfnamefont {M.}~\bibnamefont
  {Barkeshli}},\ }\href@noop {} {\bibinfo {title} {{Generalized $U(1)$ Gauge
  Field Theories and Fractal Dynamics}}} (\bibinfo {year} {2018}),\ \Eprint
  {https://arxiv.org/abs/1806.01855} {arXiv:1806.01855 [cond-mat.str-el]}
  \BibitemShut {NoStop}%
\bibitem [{\citenamefont {Gromov}(2019)}]{Gromov2019}%
  \BibitemOpen
  \bibfield  {author} {\bibinfo {author} {\bibfnamefont {A.}~\bibnamefont
  {Gromov}},\ }\bibfield  {title} {\bibinfo {title} {Towards classification of
  fracton phases: The multipole algebra},\ }\href
  {https://doi.org/10.1103/PhysRevX.9.031035} {\bibfield  {journal} {\bibinfo
  {journal} {Phys. Rev. X}\ }\textbf {\bibinfo {volume} {9}},\ \bibinfo {pages}
  {031035} (\bibinfo {year} {2019})}\BibitemShut {NoStop}%
\bibitem [{\citenamefont {Hart}\ and\ \citenamefont
  {Nandkishore}(2021)}]{Hart2021TypeII}%
  \BibitemOpen
  \bibfield  {author} {\bibinfo {author} {\bibfnamefont {O.}~\bibnamefont
  {Hart}}\ and\ \bibinfo {author} {\bibfnamefont {R.}~\bibnamefont
  {Nandkishore}},\ }\href@noop {} {\bibinfo {title} {Experimental signatures of
  gapless type ii fracton phases}} (\bibinfo {year} {2021}),\ \Eprint
  {https://arxiv.org/abs/2106.15631} {arXiv:2106.15631 [cond-mat.str-el]}
  \BibitemShut {NoStop}%
\bibitem [{\citenamefont {Schmitz}(2019)}]{Schmit}%
  \BibitemOpen
  \bibfield  {author} {\bibinfo {author} {\bibfnamefont {A.~T.}\ \bibnamefont
  {Schmitz}},\ }\href@noop {} {\bibinfo {title} {Distilling fractons from
  layered subsystem-symmetry protected phases}} (\bibinfo {year} {2019}),\
  \Eprint {https://arxiv.org/abs/1910.04765} {arXiv:1910.04765 [quant-ph]}
  \BibitemShut {NoStop}%
\bibitem [{\citenamefont {Gromov}(2020)}]{Gromov2020duality}%
  \BibitemOpen
  \bibfield  {author} {\bibinfo {author} {\bibfnamefont {A.}~\bibnamefont
  {Gromov}},\ }\href@noop {} {\bibinfo {title} {{A Duality Between U(1) Haah
  Code and 3D Smectic A Phase}}} (\bibinfo {year} {2020}),\ \Eprint
  {https://arxiv.org/abs/2002.11817} {arXiv:2002.11817 [cond-mat.str-el]}
  \BibitemShut {NoStop}%
\bibitem [{\citenamefont {Fontana}\ \emph {et~al.}(2022)\citenamefont
  {Fontana}, \citenamefont {Gomes},\ and\ \citenamefont
  {Chamon}}]{WesleiChamon2021}%
  \BibitemOpen
  \bibfield  {author} {\bibinfo {author} {\bibfnamefont {W.~B.}\ \bibnamefont
  {Fontana}}, \bibinfo {author} {\bibfnamefont {P.~R.~S.}\ \bibnamefont
  {Gomes}},\ and\ \bibinfo {author} {\bibfnamefont {C.}~\bibnamefont
  {Chamon}},\ }\bibfield  {title} {\bibinfo {title} {{Field Theories for
  type-II fractons}},\ }\href {https://doi.org/10.21468/SciPostPhys.12.2.064}
  {\bibfield  {journal} {\bibinfo  {journal} {SciPost Phys.}\ }\textbf
  {\bibinfo {volume} {12}},\ \bibinfo {pages} {64} (\bibinfo {year}
  {2022})}\BibitemShut {NoStop}%
\bibitem [{\citenamefont {Gopalakrishnan}\ and\ \citenamefont
  {Zakirov}(2018)}]{GopalakrishnanAutomata2018}%
  \BibitemOpen
  \bibfield  {author} {\bibinfo {author} {\bibfnamefont {S.}~\bibnamefont
  {Gopalakrishnan}}\ and\ \bibinfo {author} {\bibfnamefont {B.}~\bibnamefont
  {Zakirov}},\ }\bibfield  {title} {\bibinfo {title} {Facilitated quantum
  cellular automata as simple models with non-thermal eigenstates and
  dynamics},\ }\href {https://doi.org/10.1088/2058-9565/aad759} {\bibfield
  {journal} {\bibinfo  {journal} {Quantum Science and Technology}\ }\textbf
  {\bibinfo {volume} {3}},\ \bibinfo {pages} {044004} (\bibinfo {year}
  {2018})}\BibitemShut {NoStop}%
\bibitem [{\citenamefont {Alba}\ \emph {et~al.}(2019)\citenamefont {Alba},
  \citenamefont {Dubail},\ and\ \citenamefont {Medenjak}}]{AlbaRule54}%
  \BibitemOpen
  \bibfield  {author} {\bibinfo {author} {\bibfnamefont {V.}~\bibnamefont
  {Alba}}, \bibinfo {author} {\bibfnamefont {J.}~\bibnamefont {Dubail}},\ and\
  \bibinfo {author} {\bibfnamefont {M.}~\bibnamefont {Medenjak}},\ }\bibfield
  {title} {\bibinfo {title} {Operator entanglement in interacting integrable
  quantum systems: The case of the rule 54 chain},\ }\href
  {https://doi.org/10.1103/PhysRevLett.122.250603} {\bibfield  {journal}
  {\bibinfo  {journal} {Phys. Rev. Lett.}\ }\textbf {\bibinfo {volume} {122}},\
  \bibinfo {pages} {250603} (\bibinfo {year} {2019})}\BibitemShut {NoStop}%
\bibitem [{\citenamefont {Iaconis}(2021)}]{IaconisPRXComplexity}%
  \BibitemOpen
  \bibfield  {author} {\bibinfo {author} {\bibfnamefont {J.}~\bibnamefont
  {Iaconis}},\ }\bibfield  {title} {\bibinfo {title} {Quantum state complexity
  in computationally tractable quantum circuits},\ }\href
  {https://doi.org/10.1103/PRXQuantum.2.010329} {\bibfield  {journal} {\bibinfo
   {journal} {PRX Quantum}\ }\textbf {\bibinfo {volume} {2}},\ \bibinfo {pages}
  {010329} (\bibinfo {year} {2021})}\BibitemShut {NoStop}%
\bibitem [{\citenamefont {Smith}\ \emph {et~al.}(2017)\citenamefont {Smith},
  \citenamefont {Knolle}, \citenamefont {Kovrizhin},\ and\ \citenamefont
  {Moessner}}]{Smith2017DisorderFree}%
  \BibitemOpen
  \bibfield  {author} {\bibinfo {author} {\bibfnamefont {A.}~\bibnamefont
  {Smith}}, \bibinfo {author} {\bibfnamefont {J.}~\bibnamefont {Knolle}},
  \bibinfo {author} {\bibfnamefont {D.~L.}\ \bibnamefont {Kovrizhin}},\ and\
  \bibinfo {author} {\bibfnamefont {R.}~\bibnamefont {Moessner}},\ }\bibfield
  {title} {\bibinfo {title} {Disorder-free localization},\ }\href
  {https://doi.org/10.1103/PhysRevLett.118.266601} {\bibfield  {journal}
  {\bibinfo  {journal} {Phys. Rev. Lett.}\ }\textbf {\bibinfo {volume} {118}},\
  \bibinfo {pages} {266601} (\bibinfo {year} {2017})}\BibitemShut {NoStop}%
\bibitem [{\citenamefont {Smith}(2019)}]{Smith2019Thesis}%
  \BibitemOpen
  \bibfield  {author} {\bibinfo {author} {\bibfnamefont {A.}~\bibnamefont
  {Smith}},\ }\href {https://doi.org/10.1007/978-3-030-20851-6} {\emph
  {\bibinfo {title} {Disorder-free localization}}}\ (\bibinfo  {publisher}
  {Springer},\ \bibinfo {year} {2019})\BibitemShut {NoStop}%
\bibitem [{\citenamefont {Qi}\ and\ \citenamefont {Lucas}(pear)}]{marvin}%
  \BibitemOpen
  \bibfield  {author} {\bibinfo {author} {\bibfnamefont {M.}~\bibnamefont
  {Qi}}\ and\ \bibinfo {author} {\bibfnamefont {A.}~\bibnamefont {Lucas}},\
  }\href@noop {} {} (\bibinfo {year} {to appear})\BibitemShut {NoStop}%
\bibitem [{\citenamefont {Fox}(1983)}]{PhysRevA.27.3216}%
  \BibitemOpen
  \bibfield  {author} {\bibinfo {author} {\bibfnamefont {R.~F.}\ \bibnamefont
  {Fox}},\ }\bibfield  {title} {\bibinfo {title} {Long-time tails and
  diffusion},\ }\href {https://doi.org/10.1103/PhysRevA.27.3216} {\bibfield
  {journal} {\bibinfo  {journal} {Phys. Rev. A}\ }\textbf {\bibinfo {volume}
  {27}},\ \bibinfo {pages} {3216} (\bibinfo {year} {1983})}\BibitemShut
  {NoStop}%
\bibitem [{\citenamefont {Kovtun}\ and\ \citenamefont
  {Yaffe}(2003)}]{Kovtun:2003vj}%
  \BibitemOpen
  \bibfield  {author} {\bibinfo {author} {\bibfnamefont {P.}~\bibnamefont
  {Kovtun}}\ and\ \bibinfo {author} {\bibfnamefont {L.~G.}\ \bibnamefont
  {Yaffe}},\ }\bibfield  {title} {\bibinfo {title} {{Hydrodynamic fluctuations,
  long time tails, and supersymmetry}},\ }\href
  {https://doi.org/10.1103/PhysRevD.68.025007} {\bibfield  {journal} {\bibinfo
  {journal} {Phys. Rev. D}\ }\textbf {\bibinfo {volume} {68}},\ \bibinfo
  {pages} {025007} (\bibinfo {year} {2003})},\ \Eprint
  {https://arxiv.org/abs/hep-th/0303010} {arXiv:hep-th/0303010} \BibitemShut
  {NoStop}%
\bibitem [{\citenamefont {Sala}\ \emph {et~al.}(2021)\citenamefont {Sala},
  \citenamefont {Lehmann}, \citenamefont {Rakovszky},\ and\ \citenamefont
  {Pollmann}}]{sala2021dynamics}%
  \BibitemOpen
  \bibfield  {author} {\bibinfo {author} {\bibfnamefont {P.}~\bibnamefont
  {Sala}}, \bibinfo {author} {\bibfnamefont {J.}~\bibnamefont {Lehmann}},
  \bibinfo {author} {\bibfnamefont {T.}~\bibnamefont {Rakovszky}},\ and\
  \bibinfo {author} {\bibfnamefont {F.}~\bibnamefont {Pollmann}},\ }\href@noop
  {} {\bibinfo {title} {Dynamics in systems with modulated symmetries}}
  (\bibinfo {year} {2021}),\ \Eprint {https://arxiv.org/abs/2110.08302}
  {arXiv:2110.08302 [cond-mat.stat-mech]} \BibitemShut {NoStop}%
\bibitem [{\citenamefont {Abramovitz}\ and\ \citenamefont
  {Stegun}(1964)}]{abramovitz1964handbook}%
  \BibitemOpen
  \bibfield  {author} {\bibinfo {author} {\bibfnamefont {M.}~\bibnamefont
  {Abramovitz}}\ and\ \bibinfo {author} {\bibfnamefont {I.~A.}\ \bibnamefont
  {Stegun}},\ }\href@noop {} {\emph {\bibinfo {title} {Handbook of mathematical
  functions. With formulas, graphs and mathematical tables}}}\ (\bibinfo
  {publisher} {Dover},\ \bibinfo {year} {1964})\BibitemShut {NoStop}%
\end{thebibliography}%
\end{document}